\documentclass[trackchanges]{aastex62}
\usepackage{verbatim}
\shorttitle{Comparison between FDS and non-FDS Superluminal Sources}
\shortauthors{H. B. Xiao et al.}

\begin{document}
\title{Comparison between {\textit{Fermi}} Detected and {\textit{non-Fermi}} Detected Superluminal  Sources}

\correspondingauthor{Junhui Fan}
\email{fjh@gzhu.edu.cn}

\author{Hubing Xiao}
\affiliation{Center for Astrophysics, Guangzhou University \\
Guangzhou, 510006, China}
\affiliation{Department of Physics and Astronomy, University of Padova \\
Padova PD, 35131, Italy}

\author{Junhui Fan}
\affiliation{Center for Astrophysics, Guangzhou University \\
Guangzhou, 510006, China}
\affiliation{Astronomy Science and Technology Research Laboratory of Department of Education of Guangdong Province \\
Guangzhou, 510006, China}

\author{Jianghe Yang}
\affiliation{Department of Physics and Electronics Science, Hunan University of Arts and Science \\
Changde, 415000, China}

\author{Yi Liu}
\affiliation{Center for Astrophysics, Guangzhou University \\
Guangzhou, 510006, China}
\affiliation{Astronomy Science and Technology Research Laboratory of Department of Education of Guangdong Province \\
Guangzhou, 510006, China}

\author{Yuhai Yuan}
\affiliation{Center for Astrophysics, Guangzhou University \\
Guangzhou, 510006, China}
\affiliation{Astronomy Science and Technology Research Laboratory of Department of Education of Guangdong Province \\
Guangzhou, 510006, China}

\author{Jun Tao}
\affiliation{Shanghai Astronomical Observatory, Chinese Academy of Sciences \\
 Shanghai 210006, China}

\author{Denise Costantin}
\affiliation{Center for Astrophysics, Guangzhou University \\
Guangzhou, 510006, China}
\affiliation{Astronomical Science and Technology Research Laboratory, Department of Education \\
Guangdong Province, China}
\affiliation{Department of Statistical Science, University of Padova \\
Padova PD, 35131, Italy}

\author{Yutao Zhang}
\affiliation{Center for Astrophysics, Guangzhou University \\
Guangzhou, 510006, China}
\affiliation{Astronomy Science and Technology Research Laboratory of Department of Education of Guangdong Province \\
Guangzhou, 510006, China}

\author{Zhiyuan Pei}
\affiliation{Center for Astrophysics, Guangzhou University \\
Guangzhou, 510006, China}
\affiliation{Department of Physics and Astronomy, University of Padova \\
Padova PD, 35131, Italy}

\author{Lixia Zhang}
\affiliation{Center for Astrophysics, Guangzhou University \\
Guangzhou, 510006, China}
\affiliation{Astronomy Science and Technology Research Laboratory of Department of Education of Guangdong Province \\
Guangzhou, 510006, China}

\author{Wenxin Yang}
\affiliation{Center for Astrophysics, Guangzhou University \\
Guangzhou, 510006, China}
\affiliation{Astronomy Science and Technology Research Laboratory of Department of Education of Guangdong Province \\
Guangzhou, 510006, China}

\begin{abstract}
{
Active galactic nuclei (AGNs) have been attracting research attention due to
their special observable properties. Specifically, a majority of AGNs are detected by
Fermi-LAT missions, but not by Fermi-LAT, which raises the question of weather any differences exist between the two.
To answer this issue, we compile a sample of 291 superluminal AGNs (189 FDSs and 102 non-FDSs) from
available multi-wavelength
 radio, optical, and X-ray (or even $\gamma$-ray) data and
 Doppler factors and proper motion ($\mu$) (or apparent velocity ($\beta_{\rm{app}}$));
 calculated the apparent velocity from their proper motion,
 Lorentz factor ($\Gamma$),
 viewing angle ($\phi$) and co-moving viewing angle ($\phi_{co}$) for the sources
 with available Doppler factor ($\delta$);
 and performed some statistical analyses  for both types.
Our study indicated that
1. In terms of average values, FDSs have higher
        proper motions ($\mu$),
         apparent velocities ($\beta_{\rm app}$),
         Doppler factor ($\delta$),
         Lorentz factor ($\Gamma$),
and
    smaller viewing angle ($\phi$).
Nevertheless, there is no clear difference in co-moving viewing angles ($\phi_{\rm co}$).
The results reveal that that FDSs show stronger beaming effect than non-FDSs.
2. In terms of correlations:
(1) Both sources show positive, mutually correlated fluxes, which become closer in de-beamed fluxes;
(2) With respect to apparent velocities and $\gamma$-ray luminosity, there is a tendency for the brighter sources to have higher velocities;
(3) With regard to viewing angle and observed $\gamma$-ray luminosity, ${\rm log}\phi=
 -(0.23\pm0.04 ) {\rm log}L_{\rm \gamma}+( 11.14\pm 1.93 )$, while for the co-moving viewing angle and the intrinsic $\gamma$-ray luminosity, ${\rm log}\phi_{\rm co}=( 0.09 \pm 0.01 ) {\rm log}L_{\rm \gamma}^{\rm in}-( 1.73 \pm 0.48 )$. These  correlations show that the luminous $\gamma$-ray sources have smaller viewing angles and a larger co-moving viewing angle, which indicate a stronger beaming effect in $\gamma$-ray emissions.
}
\end{abstract}



\section{introduction}
The nature of AGNs is still an open question in astrophysics.
AGNs contain a broad wavelength band emission,
from radio to very high energy (VHE) band.
 Blazars, as a very extreme subclass of AGNs,
 show rapid and high variability,
 high and variable polarization,
 variable and strong $\gamma$-ray emission and even superluminal motion
 (\citealt{Fan2013a};
  \citealt{Fan2013b}).
 These extreme observational properties of blazars are due to the fact that
 they host a relativistic jet pointing to the observer
 (\citealt{Blandford1979}).
  Blazars
  have two subclasses,
  namely BL Lacertae objects (BL Lacs) and
  flat spectrum radio quasars (FSRQs).
  BL Lacs show weak or no emission lines while
  FSRQs show strong emission line features.
 The classifications of blazars based on the spectral energy distributions (SEDs) can be found in
  \citet{Padovani1995},
  \citet{Nieppola2006},
  \citet{Abdo2010a},
  \citet{Fan2016},
  \citet{Lin2018} and
  \citet{ZF2019}.

The $\gamma$-ray emissions of blazars have caught astronomers' attention to
 investigate the mechanism of the high energetic $\gamma$-ray emissions.
 There have been two generations of $\gamma$-ray experiment,
  EGRET (the Energetic Gamma-Ray Experiment Telescope, on-board the $Compton\ Gamma-Ray\ Observatory$) and
  {\textit{Fermi-LAT}} ({\textit{Fermi}} Large Area $Gamma-Ray\ Space\ Telescope$),
  which provide us good opportunities to detect strong $\gamma$-ray sources.
  Based on the observations of EGRET, correlation analyses between the $\gamma$-ray emissions
  and those at lower energy bands have been performed to study the beaming effect (
  \citealt{Dondi1995};
  \citealt{Xie1997};
  \citealt{Fan1999};
  \citealt{Cheng2000}, and reference therein).
{\textit{Fermi-LAT}}, a successor to ERGET,
 detected more than 1000 blazars (see
  \citealt{Abdo2010b};
  \citealt{Nolan2012};
  \citealt{Acero2015};
  \citealt{Ackermann2015}).
 The strong $\gamma$-ray emissions in blazars suggest the existence of
  a relativistic beaming effect,
  which is discussed in many papers (see \citealt{Arshakian2010};
  \citealt{Fan2013a};   
  \citealt{Fan2013b};
  \citealt{Fan2014};
  \citealt{Fan2017};
  \citealt{FanJi2014};
  \citealt{Giovannini2014};
  \citealt{Giroletti2012};
  \citealt{Kovalev2009};
  \citealt{Massaro2013a}; 
  \citealt{Massaro2013b};
  \citealt{Pushkarev2010};
  \citealt{Savolainen2010};
  \citealt{Xiao2015};
  \citealt{Pei2016};
  \citealt{Yang2017};
  \citealt{Yang2018a};
  \citealt{ZF2018}).
The $\gamma$-ray emissions are also used to estimate the beaming boosting factors (Doppler factors) for some $\gamma$-ray loud sources
(\citealt{Mattox1993};
\citealt{Dondi1995};
\citealt{vonMontigny1995};
\citealt{Cheng1999};
\citealt{Fan1999};
\citealt{Fan2013a};
\citealt{Fan2013b};
\citealt{Fan2014};
\citealt{Fan2005}).
\citet{Yang2018b} studied the effective spectral index properties, then suggested that synchrotron self-Compton (SSC) model could explain the main process for highly energetic $\gamma$ rays in BL Lacs. 
Moreover, the DArk Matter Particle Explorer (DAMPE),
was successfully launched into a sun-synchronous orbit at the altitude of 500 km
on 2015 December 17th from the Jiuquan launch base.
DAMPE offers a new opportunity for advancing our knowledge of cosmic rays,
dark matter,
and gamma-ray astronomy as well (\citealt{Chang2017}).
This marks a new generation of astrophysics,
which has bound particles physics and astronomy together tightly.

For blazars,
 the Doppler factor ($\delta = [\Gamma (1- \beta \rm{cos} \phi)]^{-1}$) is an important parameter,
 where $\Gamma = (1 - \beta ^{2})^{1/2}$ is a bulk Lorentz factor,
 $\beta$ is the jet speed in units of the speed of light,
 and $\phi$ is a viewing angle between the jet and the line-of-sight.
 The Doppler factor is a key quantity in jets since it determines how
 much flux densities are boosted and timescales compressed in the observer frame.
The Doppler factor,
 although a crucial parameter in the blazar paradigm dictating all of the observed properties of blazars,
 is very difficult to estimate since there is no direct determining method for either $\beta$ or $\phi$.
For this reason,
 many indirect methods have been proposed in order to estimate $\delta$,
  which usually involves different energetic
  (e.g., 
  \citealt{Ghisellini1993};
  \citealt{Mattox1993};
  \citealt{Fan2013a};
  \citealt{Fan2014})
  and/or causality arguments
   (\citealt{Lahteenimaki1999};
    \citealt{Hovatta2009};
    \citealt{Jorstad2005};
    \citealt{Jorstad2017};
     \citealt{Liodakis2018}) or
 fitting the spectral energy distribution (SED,
   \citealt{Ghisellini2014};
   \citealt{Zhang2012};
   \citealt{Chen2018}) of $\gamma$-ray emitting blazars.
  \citet{Lahteenimaki1999} proposed to estimate the Doppler factor
  ($\delta_{\rm  var}$) using radio flux density variations.
  They obtained the timescales for radio emissions,
  assumed the timescales to represent the emission size,
  and got a brightness temperature ($T_{\rm B}^{\rm ob}$).
  If the intrinsic brightness temperature is assumed to be
$T_{\rm B}^{\rm in}=5\times 10^{10} \rm{K}$
 and the difference between the two brightness temperatures is from the beaming effect,
 then a variation Doppler factor,
 $\delta_{\rm var} = (T_{\rm B}^{\rm ob}/T_{\rm B}^{\rm in})^{1/3}$ can be estimated.
 This method was used to estimate a large sample with longer coverage of radio observations
 (see \citealt{Hovatta2009}).
  \citet{Fan2009} and
  \citet{Savolainen2010} also adopted that method to estimate the Doppler factor in the radio band.
 Furthermore, the Doppler factor was also estimated for the $\gamma$-ray loud blazars,
  \citet{Fan2013a} and \citet{Fan2014} suggested a Doppler factor  can be expressed as
$$\delta \geq[1.54 \times 10^{-3} (1+z)^{4+2\alpha} (\frac{d_{\rm L}}{ \rm{Mpc}})^{2} (\frac{\Delta T}{\rm{hr}})^{-1} (\frac{F_{\rm{1 KeV}}}{\rm{\mu Jy}}) (\frac{E_{\rm{\gamma}}}{\rm{GeV}})^{\alpha}]^{\frac{1}{4+2\alpha}}$$
here $\Delta T$ is the time scale in units of hour,
$\alpha$ is the X-ray spectral index,
$F_{\rm{1 KeV}}$ is the flux density at 1 KeV in units of $\rm{\mu Jy}$,
$E_{\rm{\gamma}}$ is the energy in units of GeV,
 at which the $\gamma$-rays are detected,
 and $d_{\rm L}$ is luminosity distance in units of Mpc.

Superluminal motion is also an interesting observational property of blazars.
Thanks to the very large baseline interferometry (VLBI) established by Europe, Canada, United States, Russia and so on,
with high angular resolution at milliarcsecond,
many AGNs show interesting observational results,
some compact radio sources consist of more than one component,
and some of these components seem to be separating at apparent velocities being greater than the speed of light.
A parameter, $\beta_{\rm app}(= v / c)$,
is introduced to value the apparent velocity.
If $\beta_{\rm app} > 1$,
 then it is called to be superluminal,
 and this kind of sources are called superluminal sources.

The first apparent superluminal motion was observed from 3C 279,
a component moving away from the quasar core at nearly ten times the speed of light was detected.
 \citet{VC1994} listed 66 extragalactic sources,
from which they found multi-epoch VLBI internal proper motions,
then investigated several modifications to a simple relativistic beam concept
and its statistical effects on apparent velocity.
They also checked the distribution of $\beta_{\rm app}$ for lobe-selected and
core-selected quasars respectively and obtained the $\beta_{\rm app}$ for different
object categories to be in general agreement with an AGN unification model.
In 1996, \citeauthor{Fan1996} compiled a sample of 48 superluminal motion sources to investigate the beaming effect,
 found that the core dominance is an indicator of the orientation of the emission,
 and proposed that the superluminal motion and beaming effect are probably the same things.
 \citet{Kellermann2003} presented a sample of 96 superluminal sources to
 study the nature of the relativistic beaming effect in blazars and
 their surrounding environment of the massive black holes.
 They found that most of the blazars show an outward flow away from the centre core
 while a few sources show the opposite direction of features,
 and there is no simple correlation between timescale of flux changes and apparent velocities.
 In 2008, \citeauthor{ZF2008} collected an up-to-date sample of 123 superluminal sources
  including 84 quasars,
   27 BL Lac objects and 12 galaxies,
   calculated the apparent velocities for each source,
   and found that the radio emissions are strongly boosted by the beaming effect and
   the superluminal motion is the same thing as the beaming effect in AGNs.

For the details of kinematics in superluminal motion,
 \citet{Britzen2008} presented a detailed kinematic analysis of
 the complete flux-density limited Caltech-Jodrell Bank Flat-spectrum
 (named CJF \footnote{http://www.mpifr-bonn.mpg.de/staff/sbritzen/cjf.html}) sources.
 CJF survey computed 2D kinematic models based on the optimal model-fitting parameters
 of multi-epoch VLBA observations,
 then investigated possible correlations between the apparent proper motions
 and some other parameters in AGN jets.
 They found a strong correlation between the 5 GHz luminosity and apparent velocity.
 Based on the data of  MOJAVE \footnote{http://www.physics.purdue.edu/MOJAVE/}
 (Monitoring of Jets in Active galactic nuclei with VLBA Experiments) sample,
  \citet{Lister2009} discussed the jet kinematics of a complete flux-density-limited sample of
 135 radio-loud AGNs resulting from a 13-years program,
 investigated the structure and evolution of parsec-scale jet phenomena,
 and found there is an overwhelming tendency to display outward motions,
 only eight inward moving components.
 \citet{Lister2013} studied 200 AGNs parsec-scale jet orientation variations and superluminal motion,
found a general trend of increasing apparent speed with distance down the jet
for both radio galaxies and BL Lac objects.

Since {\textit{Fermi-LAT}} was launched in 2008, the 4-year catalogue includes 1591 AGNs.
Although AGNs are the main detection result by {\textit{Fermi-LAT}}, while there are many AGNs are not detected
by Fermi. Why are some AGNs detected by {\textit{Fermi-LAT}} and others are not? To answer this question,
 we compile a large superluminal sample (189 FDSs and 102 non-FDSs) and
use them to make a comparison between the FDS and the non-FDS sources and do some statistical analyses.
This work is arranged as follows:
 In section 1, we introduce our superluminal sample,
 in section 2, we will give results,
and discussions and conclusions are presented in sections 3 and  4.

Through this paper,
a $\Lambda$-CDM model with $\Omega_{\rm \Lambda} \simeq 0.7$,
$\Omega_{\rm M} \simeq 0.3$ and $\Omega_{\rm K} \simeq 0.0$,
and $H_{\rm 0} = 73 ~{\rm{km \, s^{-1}}} {\rm{Mpc^{-1}}}$ is adopted.

\section{Samples and Results}
\subsection{Samples}
From the available literature,
we compile 291 sources with superluminal motions,
including 189 ( 142 FSRQs, 39 BL Lacs, 5 galaxies and 2 uncertain type blazar candidates (BCU) and 1 unknown type of AGN  without a known redshift )
{\textit{Fermi}} detected superluminal sources (FDS) and
102 ( 98 FSRQs, 1 BL Lac and 12 galaxies and 1 unknown type of AGN without a known redshift )
{\textit{non-Fermi}} detected superluminal (non-FDS) sources,
where {\textit{Fermi}} detected sources mean these sources are detected by {\textit{Fermi-LAT}} telescope and listed in the {\textit{Fermi}} AGN catalogues.

There are 816 components for the 189 FDS sources in total,
30 of them have just one component.
In the present sample,
 we also include the $\gamma$-ray emission source,
 0007+106 (III ZW 2),
 which was classified as an $\gamma$-ray source by \citet{Liao2016}.
 All the FDS sources are listed in Table \ref{Tab-FDS-simp}.

\begin{deluxetable}{c|c|c|c|c|c|c|c|c|c|c|c|c|c|c}
\tabletypesize{\scriptsize}
 \rotate
  \tablecaption{Superluminal Sources detected by {\textit{Fermi-LAT}}}
  \tablewidth{0pt}
 \tablehead{
  \colhead{ FGL name }&
  \colhead{ Class } &
  \colhead{ redshift } &
  \colhead{ $\delta_R$ } &
  \colhead{ Ref } &
  \colhead{ $m_o$ } &
  \colhead{ Ext  }&
  \colhead{ S$_R$ }&
  \colhead{ S$_X$ }&
  \colhead{ $\Gamma_{\gamma}$ }&
  \colhead{ F$\gamma$ }&
  \colhead{ $\mu$ }&
  \colhead{ comp }&
  \colhead{ Ref } &
  \colhead{ $\beta$ }\\
  \colhead{ Other name }&
  \colhead{ }&
  \colhead{ }&
  \colhead{ }&
  \colhead{ }&
  \colhead{ $magnitude $ }&
  \colhead{ }&
  \colhead{ $mJy$ }&
  \colhead{ $1 \times 10^{-12} ~cgs$ }&
  \colhead{ }&
  \colhead{ $ph/cm^{2}/s$ }&
  \colhead{ $\mu as/yr$ }&
  \colhead{ }&
  \colhead{ }&
  \colhead{ }\\
  \colhead{(1) }&
  \colhead{(2) }&
  \colhead{(3) }&
  \colhead{(4) }&
  \colhead{(5) }&
  \colhead{(6) }&
  \colhead{(7) }&
  \colhead{(8) }&
  \colhead{(9) }&
  \colhead{(10) }&
  \colhead{(11) }&
  \colhead{(12) }&
  \colhead{(13) }&
  \colhead{(14) }&
  \colhead{(15) }
    }
 \startdata
0007+106 	&	G	&	0.089	&	2.51	&	L18	&	15.8	&	0.227	&	98	&	6.14	&		&		&	204 $\pm$ 12	&	1	&	MOJAVE	&	1.197$\pm$0.069	\\	
III ZW 2	&		&		&		&		&		&		&		&		&		&		&	269 $\pm$ 50	&	4	&	MOJAVE	&	1.58$\pm$0.29	\\	\hline
1FGL J1159.4-2149	&	F	&	0.927	&		&		&	17.8	&	0.104	&	386	&	0	&		&		&	9.3 $\pm$ 3.8	&	1	&	MOJAVE	&	0.46$\pm$0.19	\\	
1157-215 	&		&		&		&		&		&		&		&		&		&		&	83 $\pm$ 22	&	2	&	MOJAVE	&	4.1$\pm$1.1	\\	
	&		&		&		&		&		&		&		&		&		&		&	26.2 $\pm$ 8.0	&	3	&	MOJAVE	&	1.30$\pm$0.40	\\	 \hline
1FGL J1245.8-0632	&	F	&	1.286	&		&		&	19.6	&	0.068	&	551	&	0	&		&		&	358 $\pm$ 53	&	1	&	MOJAVE	&	22.5$\pm$3.4	\\	
1243-072 	&		&		&		&		&		&		&		&		&		&		&	17.6 $\pm$ 8.9e	&	3	&	MOJAVE	&	1.11$\pm$0.56	\\	
	&		&		&		&		&		&		&		&		&		&		&	40 $\pm$ 16	&	4	&	MOJAVE	&	2.51$\pm$1.00	\\	 \hline
2FGLJ2148.2+0659 	&	F	&	0.999	&	15.6	&	H09	&	15.9	&	0.186	&	2590	&	1.46	&	2.77	&	3.90E-10	&	59.0 $\pm$ 1.8	&	2a 	&	MOJAVE	&	3.092$\pm$0.096	\\	
2145+067	&		&		&		&		&		&		&		&		&		&		&	50.1 $\pm$ 9.1	&	3a 	&	MOJAVE	&	2.63$\pm$0.48	\\	
	&		&		&		&		&		&		&		&		&		&		&	49.2 $\pm$ 3.3	&	5	&	MOJAVE	&	2.58$\pm$0.17	\\	
	&		&		&		&		&		&		&		&		&		&		&	59.3 $\pm$ 3.8	&	7	&	MOJAVE	&	3.11$\pm$0.20	\\	
	&		&		&		&		&		&		&		&		&		&		&	27.6 $\pm$ 3.8	&	8	&	MOJAVE	&	1.45$\pm$0.20	\\	 \hline
3FGL J0006.4+3825	&	F	&	0.229	&		&		&	17.6	&	0.205	&	572	&	0.75	&	2.617	&	6.06E-10	&	9$\pm$38	&	C1	&	CJF	&	0.12$\pm$0.52	\\	
0003+380	&		&		&		&		&		&		&		&		&		&		&	135$\pm$37	&	C2	&	CJF	&	1.85$\pm$0.51	\\	
	&		&		&		&		&		&		&		&		&		&		&	145	&	C3	&	CJF	&	1.99$\pm$0	\\	
	&		&		&		&		&		&		&		&		&		&		&	336	&	C4	&	CJF	&	4.6$\pm$0	\\	 \hline
\enddata
\label{Tab-FDS-simp}
\tablecomments{Only five objects were shown here and the whole Table will be given in the electronic version.
column (1) gives the {\textit{Fermi}} name (other name),
column (2) classification, F stands for FSRQs, B for BL Lacs, Sy for Seyfert galaxies, Un for unknown type AGNs,
column (3) redshift,
column (4) Doppler factor,
column (5) reference for Doppler factor,
column (6) apparent magnitude ($R_{\rm{mag}}$) from BZCAT \footnote{http://www.asdc.asi.it/bzcat/} (Massaro et al. 2009),
column (7) Galactic extinction ($A_{\rm R}$) from NED,
column (8) flux density at 1.4 GHz from BZCAT,
column (9) X-ray flux in the 0.1-2.4 KeV band from BZCAT,
column (10) $\gamma$-ray photon spectral index,
column (11) $\gamma$-ray photon flux arrange 1-100 GeV (Acero et al. 2015),
column (12) proper motion $\mu$ in microarcsecond per year,
column (13) components for proper motion,
column (14) reference for proper motion,
column (15) apparent velocity, $\beta_{\rm app}$.
}
\end{deluxetable}

For the 102 non-FDS sources ( 88 FSRQs, 1 BL Lac, 12 galaxies and 1 unknown type of AGN) with 400 components totally, 17 of them have just one component, they are in Table \ref{Tab-non-FDS-simp},

\begin{deluxetable}{c|c|c|c|c|c|c|c|c|c|c|c|c}
\tabletypesize{\scriptsize}
 \rotate
  \tablecaption{Superluminal Sources not detected by {\textit{Fermi-LAT}}}
  \tablewidth{0pt}
 \tablehead{
  \colhead{ FGL name }&
  \colhead{ Class } &
  \colhead{ redshift } &
  \colhead{ $\delta_R$ } &
  \colhead{ Ref } &
  \colhead{ $m_o$ } &
  \colhead{ Ext  }&
  \colhead{ S$_R$ }&
  \colhead{ S$_X$ }&
  \colhead{ $\mu$ }&
  \colhead{ comp }&
  \colhead{ Ref } &
  \colhead{ $\beta$ }\\
  \colhead{ Other name }&
  \colhead{ }&
  \colhead{ }&
  \colhead{ }&
  \colhead{ }&
  \colhead{ $magnitude$ }&
  \colhead{ }&
  \colhead{ $mJy$ }&
  \colhead{ $1 \times 10^{-12} ~cgs$ }&
  \colhead{ $\mu as/yr$ }&
  \colhead{ }&
  \colhead{ }&
  \colhead{ }\\
  \colhead{(1) }&
  \colhead{(2) }&
  \colhead{(3) }&
  \colhead{(4) }&
  \colhead{(5) }&
  \colhead{(6) }&
  \colhead{(7) }&
  \colhead{(8) }&
  \colhead{(9) }&
  \colhead{(10) }&
  \colhead{(11) }&
  \colhead{(12) }&
  \colhead{(13) }
    }
 \startdata
0003-066	&	B	&	0.347	&	5.1	&	H09	&	17.9	&	0.087	&	2051	&	0.82	&	191 $\pm$ 15	&	2	&	MOJAVE	&	4.09$\pm$0.33	\\	
	&		&		&		&		&		&		&		&		&	250 $\pm$ 39	&	3	&	MOJAVE	&	5.36$\pm$0.83	\\	
	&		&		&		&		&		&		&		&		&	50.4 $\pm$ 5.3	&	4a 	&	MOJAVE	&	1.08$\pm$0.11	\\	
	&		&		&		&		&		&		&		&		&	100 $\pm$ 16	&	5	&	MOJAVE	&	2.15$\pm$0.35	\\	
	&		&		&		&		&		&		&		&		&	54 $\pm$ 11	&	6a 	&	MOJAVE	&	1.16$\pm$0.24	\\	
	&		&		&		&		&		&		&		&		&	330.4 $\pm$ 9.8	&	8a 	&	MOJAVE	&	7.08$\pm$0.21	\\	
	&		&		&		&		&		&		&		&		&	287 $\pm$ 25	&	9	&	MOJAVE	&	6.14$\pm$0.53	\\	
	&		&		&		&		&		&		&		&		&	116 $\pm$ 23	&	14	&	MOJAVE	&	2.48$\pm$0.50	\\	\hline
0010+405	&	F	&	0.255	&		&		&		&		&		&		&	428 $\pm$ 40	&	1	&	MOJAVE	&	6.92$\pm$0.64	\\	
	&		&		&		&		&		&		&		&		&	2 $\pm$ 16e	&	2	&	MOJAVE	&	0.04$\pm$0.26	\\	
	&		&		&		&		&		&		&		&		&	2.9 $\pm$ 4.3e	&	3	&	MOJAVE	&	0.047$\pm$0.070	\\	
	&		&		&		&		&		&		&		&		&	1.1 $\pm$ 3.1e	&	4	&	MOJAVE	&	0.018$\pm$0.050	\\	\hline
0014+813	&	F	&	3.366	&		&		&	15.9	&	0.425	&	693	&	0.77	&	4$\pm$12	&	C1	&	B08	&	0.39$\pm$1.18	\\	
	&		&		&		&		&		&		&		&		&	86$\pm$15	&	C2	&	B08	&	8.49$\pm$1.48	\\	
	&		&		&		&		&		&		&		&		&	111$\pm$18	&	C3	&	B08	&	10.95$\pm$1.78	\\	\hline
0016+731	&	F	&	1.781	&	7.9	&	H09	&	18.2	&	0.735	&	1136	&	0.11	&	106.2 $\pm$ 4.4	&	1a 	&	MOJAVE	&	8.23$\pm$0.34	\\	\hline
0022+390	&	F	&	1.946	&		&		&		&		&		&		&	113$\pm$43	&	C1	&	B08	&	8.51$\pm$3.24	\\	
	&		&		&		&		&		&		&		&		&	71$\pm$57	&	C2	&	B08	&	5.35$\pm$4.29	\\	\hline
\enddata
\label{Tab-non-FDS-simp}
\tablecomments{Only five objects were shown here and the whole Table will be given in the electronic version. column (1) name,
column (2) classification, F stand for FSRQs, B for BL Lacs, Sy for Seyfert galaxies, G for galaxies,
column (3) redshift,
column (4) Doppler factor,
column (5) reference for Doppler factor,
column (6) apparent magnitude ($R_{\rm mag}$) from BZCAT,
column (7) Galactic extinction ($A_{\rm R}$) from NED,
column (8) flux at 1.4 GHz from BZCAT,
column (9) X-ray flux in the 0.1-2.4 KeV band from BZCAT,
column (10) proper motion $\mu$ microarcsecond per year,
column (11) components for proper motion,
column (12) reference for proper motion,
column (13) apparent velocity, $\beta_{\rm app}$.
}
\end{deluxetable}

\subsection{Results}
\subsubsection{Proper Motion and Apparent Velocity Distribution}

 For a proper motion ($\mu$), an apparent velocity ($\beta_{\rm app}$) can be computed by,
\begin{center}
 \begin{equation}
 \beta_{\rm app} = {\frac {\mu} {H_{\rm 0}} {\int_1^{1+z} {\frac{1} {\sqrt{ {\Omega_{\rm M}} x^3+1-{\Omega_{\rm M}} } } }dx}}.
 \end{equation}
\end{center}

\begin{figure}
 \centering
  \includegraphics[width=6in]{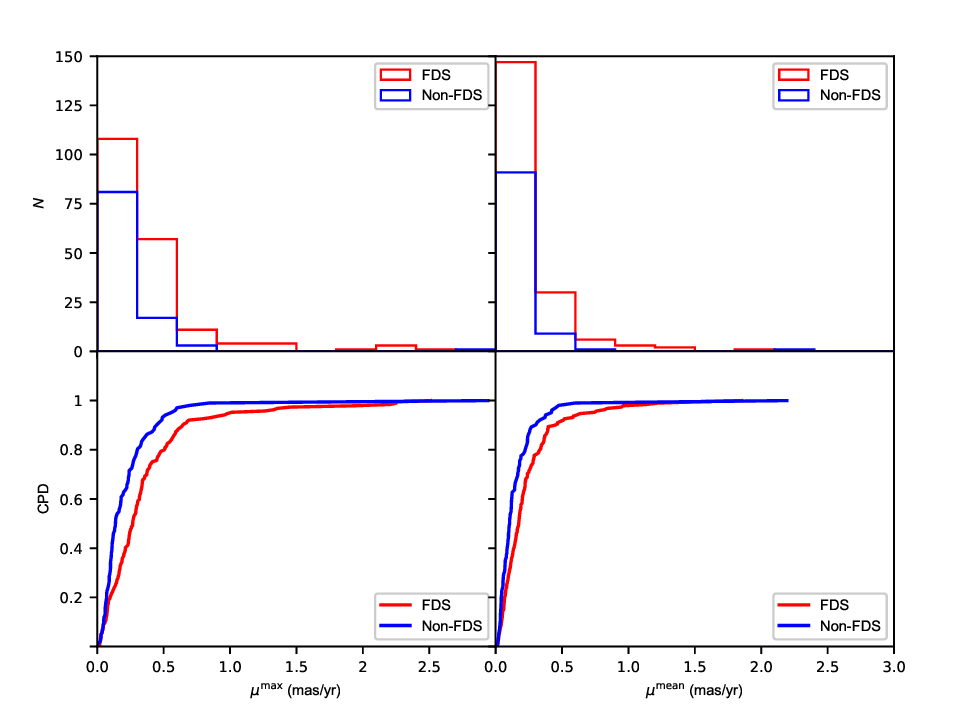}
  \caption{The distribution of maximum (left panel) / mean (right panel) proper motion ($\mu^{\rm max}$ / $\mu^{\rm mean}$ (mas/yr)). The upper panel figures are their histograms of FDS and non-FDS sources, the lower panel figures are their cumulative probability distributions (CPD).}
  \label{mu-dis}
\end{figure}

\begin{figure}
 \centering
  \includegraphics[width=6in]{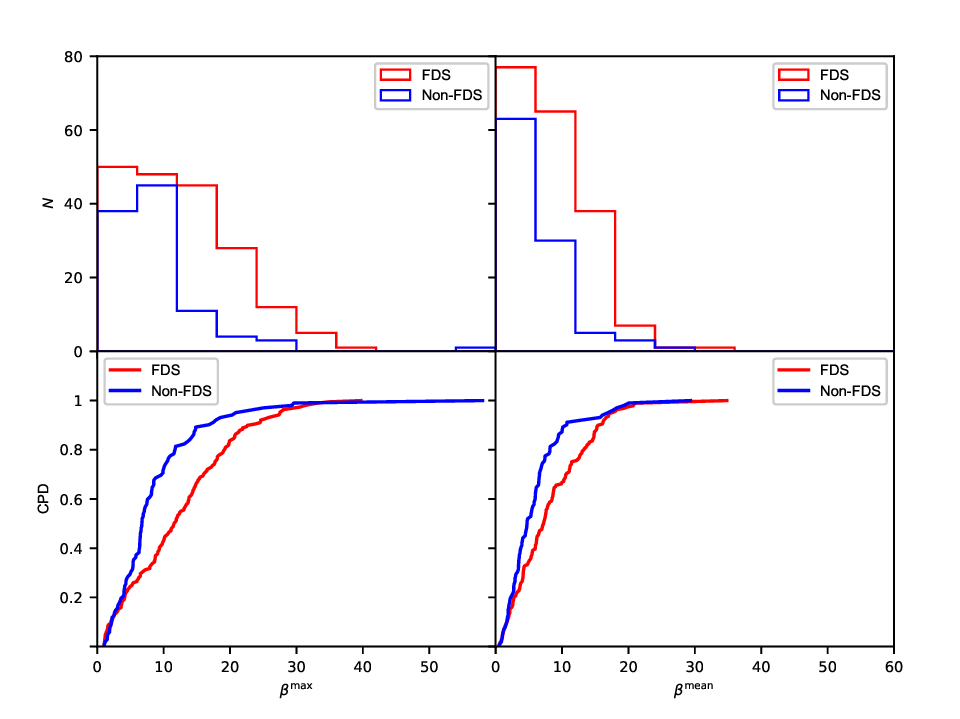}
  \caption{The distribution of maximum (left panel) / mean (right panel)  apparent velocity ($\beta_{\rm app}^{\rm max}$ / ($\beta_{\rm app}^{\rm mean}$). The upper panel figures are their histograms of FDS and non-FDS sources, the lower panel figures are their cumulative probability distributions (CPD).}
  \label{beta-dis}
\end{figure}

Hence, we calculate apparent velocities from its given proper motions if its apparent velocities are not given by MOJAVE 
(\citealt{Lister2013}; \citealt{Lister2016})
 or other reference literature for the sources in Tables \ref{Tab-FDS-simp} and \ref{Tab-non-FDS-simp}. Then we compare their maximum proper motion ($\mu^{\rm max}$) and maximum apparent velocity ($\beta^{\rm max}_{\rm app}$), averaged proper motion ($\mu^{\rm mean}$) and averaged apparent velocity ($\beta^{\rm mean}_{\rm app}$) between  FDS and non-FDS sources. The corresponding uncertainty of averaged value is expressed by an error-transmission format:
$$\sigma = \sqrt{ \sum_{i=1}^{n} (\frac{\partial f}{\partial x_{i} } )^{2} \sigma_{x_{i}}^{2}}$$
The distributions of the maximum and averaged proper motion and the apparent velocity for FDS and non-FDS sources are shown in Figures \ref{mu-dis}-\ref{beta-dis}, where the red stepped line is for FDS, and the blue stepped line for non-FDS sources.

It is found that the maximum proper motion is distributed from 0.018 to 2.510 with a mean value of $\langle \mu^{\rm max}_{\rm FDS} \rangle = 0.361 \pm 0.037 \ {\rm mas} \cdot {\rm yr^{-1}}$ for the FDS sources, and from 0.021 to 2.941 with $\langle \mu^{\rm max}_{\rm non-FDS} \rangle = 0.224 \pm 0.027 \ {\rm mas} \cdot {\rm yr^{-1}}$ for the non-FDS sources. When a Kolmogorov-Smirnov (K-S) test is adopted to the two distributions, the probability for the two distributions to come from the same distribution is $p=6.2 \times 10^{-5}$, see Figure \ref{mu-dis}.

The mean proper motion is distributed from 0.011 to 1.853 with $\langle \mu^{\rm mean}_{\rm FDS} \rangle = 0.233 \pm 0.021 \ {\rm mas} \cdot {\rm yr^{-1}}$ for the FDS sources and from 0.015 to 2.193 with $\langle \mu^{\rm mean}_{\rm non-FDS} \rangle = 0.158 \pm 0.015 \ {\rm mas} \cdot {\rm yr^{-1}}$ for the non-FDS sources, and $p=1.6 \times 10^{-4}$, see Figure \ref{mu-dis}.

From the calculations, it is found that the maximum apparent velocity $\beta^{\rm max}_{\rm app}$ is distributed from 1.04 to 39.70 with $\langle \beta_{\rm FDS}^{\rm max} \rangle = 12.36 \pm 1.64$ for the FDS sources and from 1.08 to 58.04 with $\langle \beta_{\rm non-FDS}^{\rm max} \rangle = 8.75 \pm 1.32$ for non-FDS sources, and  $p=3.3 \times10^{-7}$, see Figure \ref{beta-dis}.

For the mean apparent velocity, $\beta^{\rm mean}_{\rm app}$ is distributed from 0.53 to 34.80 with $\langle \beta_{\rm FDS}^{\rm mean} \rangle = 8.17 \pm 0.94$ for the FDS sources and  from 0.61 to 29.33 with $\langle \beta_{\rm non-FDS}^{\rm mean} \rangle = 5.99 \pm 0.78$ for non-FDSs,
and $p = 1.4 \times 10^{-4}$, see Figure \ref{beta-dis}.

\subsubsection{Correlations between Proper Motion and Redshift}
\begin{figure}
 \centering
  \includegraphics[width=5in]{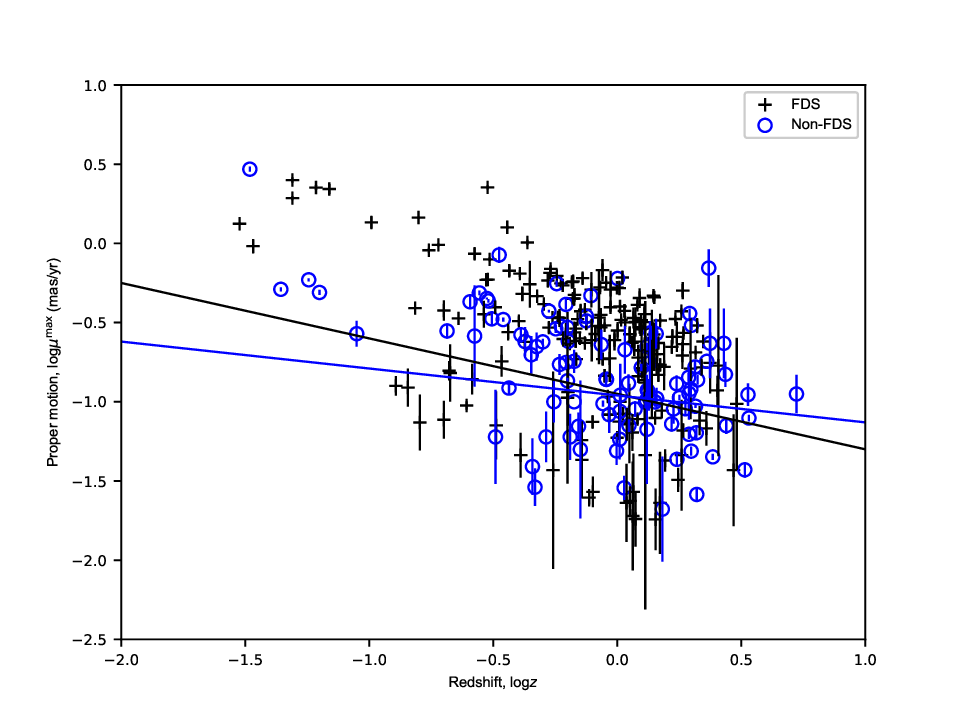}
  \caption{The plot of the maximum proper motion (log$\mu^{\rm max}$) against redshift (log$z$). The black plus stands for FDS sources and blue circle stands for non-FDS sources, black solid line and blue solid line stand for the best fitting results for FDS sources and non-FDS.}
  \label{mu-z}
\end{figure}

From the data listed in Tables \ref{Tab-FDS-simp} and \ref{Tab-non-FDS-simp}, when a linear regression fitting is adopted to the proper motion and redshift, we have
$${\rm log} \mu^{\rm max}_{\rm FDS} = -(0.35 \pm 0.10) {\rm log} z - (0.95 \pm 0.03)$$
with a correlation coefficient $r = -0.26$ and a chance probability of $p = 4.0 \times 10^{-4}$ for the FDSs, and
$${\rm log} \mu^{\rm max}_{\rm non-FDS} = -(0.14 \pm 0.11) {\rm log} z - (0.96 \pm 0.04)$$
with $r = -0.13$ and $p = 20 \%$ for the non-FDSs, see Figure \ref{mu-z}.

\subsubsection{Correlations between Apparent Velocity and Redshift}
\begin{figure}
 \centering
  \includegraphics[width=6in]{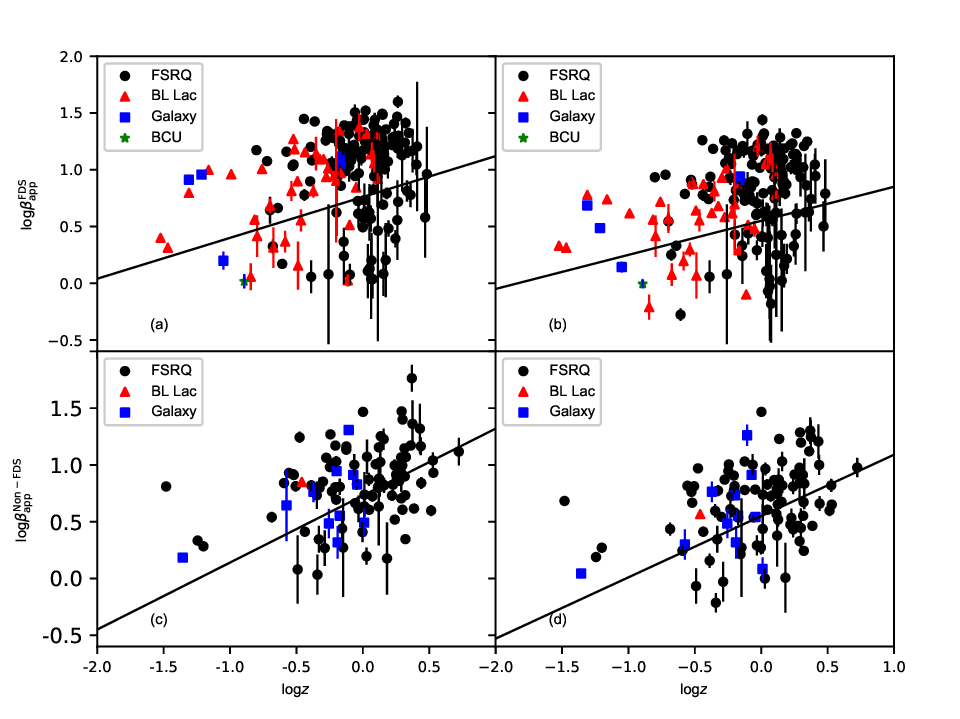}
  \caption{The plot of apparent velocity (log$\beta_{\rm app}$) against redshift (log$z$) for FDS and non-FDS sources. The dot stands for FSRQs, triangle for BL Lacs and square for Galaxies including Seyfert galaxies and normal galaxies, (a): maximum apparent velocity against redshift of FDS, (b): mean apparent velocity against redshift of FDS, and (c): maximum apparent velocity against redshift of non-FDS, (d): mean apparent velocity against redshift of non-FDS. Black solid lines in this figure represent its corresponding best linear fitting results.}
  \label{beta-z}
\end{figure}

For the apparent velocity ($\beta_{\rm app}$), we have following results for the maximum and average values of apparent velocity ($\beta_{\rm app}$),
$${\rm log} \beta^{\rm max}_{\rm FDS} = (0.36 \pm 0.09) {\rm log} z +(0.76 \pm 0.03), ~r= 0.28~{\rm and}~p= 1.4 \times 10^{-4} , ~{\rm and}$$
$${\rm log} \beta^{\rm mean}_{\rm FDS} = (0.30 \pm 0.09) {\rm log} z +(0.55 \pm 0.03), ~r= 0.24~{\rm and}~p= 1.0 \times 10^{-3}, $$
for the 186 FDSs (3 excluded sources without redshift from NED ), and \\
$${\rm log} \beta^{\rm max}_{\rm non-FDS} = (0.59 \pm 0.10) {\rm log} z +(0.73 \pm 0.03), ~r= 0.50~{\rm and}~p= 1.6 \times 10^{-7}, ~{\rm and}$$
$${\rm log} \beta^{\rm mean}_{\rm non-FDS} = (0.54 \pm 0.11) {\rm log} z +(0.55 \pm 0.03), ~r= 0.46~{\rm and}~p= 2.8 \times 10^{-6}, $$
for the 101 non-FDS as shown in Figure \ref{beta-z}.

\subsubsection{Correlation between Apparent Velocity and $\gamma$-Ray Luminosity}
For the $\gamma$-ray sources,
 the integral flux ($f$) in units of $\rm GeV \cdot cm^{-2}\cdot s^{-1}$,
 can be expressed in the form (\citealt{Fan2013b})
$$f={N_{(E_{\rm L}\sim E_{\rm U})}({\frac{1}{E_{\rm L}}-\frac{1}{E_{\rm U}}}){\rm ln}{\frac{E_{\rm U}}{E_{\rm L}}}, ~{\rm if}~ \alpha_{\rm ph} = 2,~{\rm otherwise}}$$
\begin{equation}
{f={N_{(E_{\rm L}\sim E_{\rm U})}\frac{1-\alpha_{\rm ph}}{2-\alpha_{\rm ph}}\frac{(E_{{\rm U}}^{2-\alpha_{\rm ph}}-E_{{\rm L}}^{2-\alpha_{\rm ph}})}{(E_{\rm U}^{1-\alpha_{\rm ph}}-E_{\rm L}^{1-\alpha_{\rm ph}})}}}
\end{equation}
here ${N_{(E_{\rm L}\sim E_{\rm U})}}$ is the  integral photons in the energy range of $E_{\rm L}$ and $E_{\rm U}$.
In this work, $E_{\rm L}$ and $E_{\rm U}$ are corresponding to 1 GeV and 100 GeV respectively.
Then, we calculate the $\gamma$-ray luminosity ($L_{\gamma}$) in units of $\rm erg\cdot s^{-1}$ by
\begin{equation}
{L_{\rm \gamma}=4\pi d^{2}_{\rm L}(1+z)^{\alpha_{\rm ph}-2}f}
\end{equation}
here, $L_{\rm \gamma}$ is the $\gamma$-ray luminosity, $d_{\rm L} = \frac{c(1+z)}{H_{\rm 0}}\int^{1+z}_{1}\frac{1}{\sqrt{\Omega_{\rm M}x^{3}+1-\Omega_{\rm M}}} dx$ is a luminosity distance, $(1+z)^{(\alpha_{\rm{ph}}-2)}$ stands for a K-correction, $\alpha_{\rm{ph}}$ for $\gamma$-ray photon spectral index.

\begin{figure}
 \centering
 \includegraphics[width=5in]{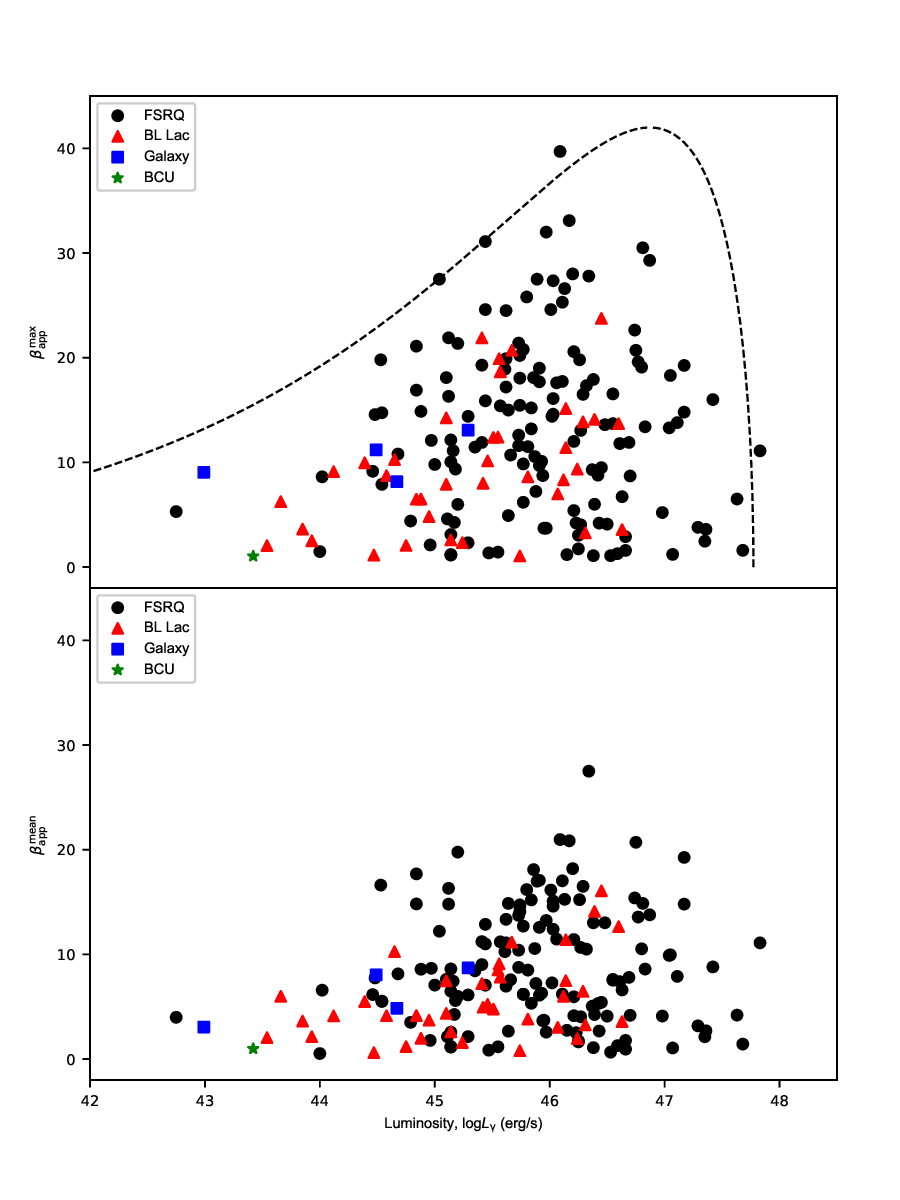}
  \caption{The plot of apparent velocity ($\beta_{\rm app}$) against the logarithmic $\gamma$-ray luminosity (log$L_{\rm \gamma}$) for FDS sources. The dot stands for FSRQs, triangle for BL Lacs and square for galaxies. The dashed curve represents an envelope with fixed $\Gamma=42$ and ${\rm log} L_{\rm in}=42$.}
  \label{beta-L}
 \end{figure}

Figure \ref{beta-L} shows the maximum and average apparent velocity against the $\gamma$-ray luminosity. The dash curved upper envelope is described by \citet{Cohen2007}, the upper envelope of this distribution traces out a single source of a given bulk Lorentz factor and intrinsic luminosity in the ($L$, $\beta_{\rm app}$) plane as the viewing angle $\phi$ changes. Such an aspect curve is plotted in Figure \ref{beta-L} for a jet with a bulk Lorentz factor of 42 and an intrinsic luminosity of log$L_{\rm in}$= 42, assuming Doppler boosting by a factor of $\delta^{3}$, by these formulas:
$\delta = \frac{1}{\Gamma(1-\beta {\rm cos} \phi)},
\beta_{\rm app} = \frac{\beta {\rm sin} \phi}{1-\beta {\rm cos} \phi}, L=L_{\rm in}\delta^{3}.$

\subsubsection{Flux Density-Flux Density Correlations}

\begin{figure*}
 \centering
 \includegraphics[width=6in]{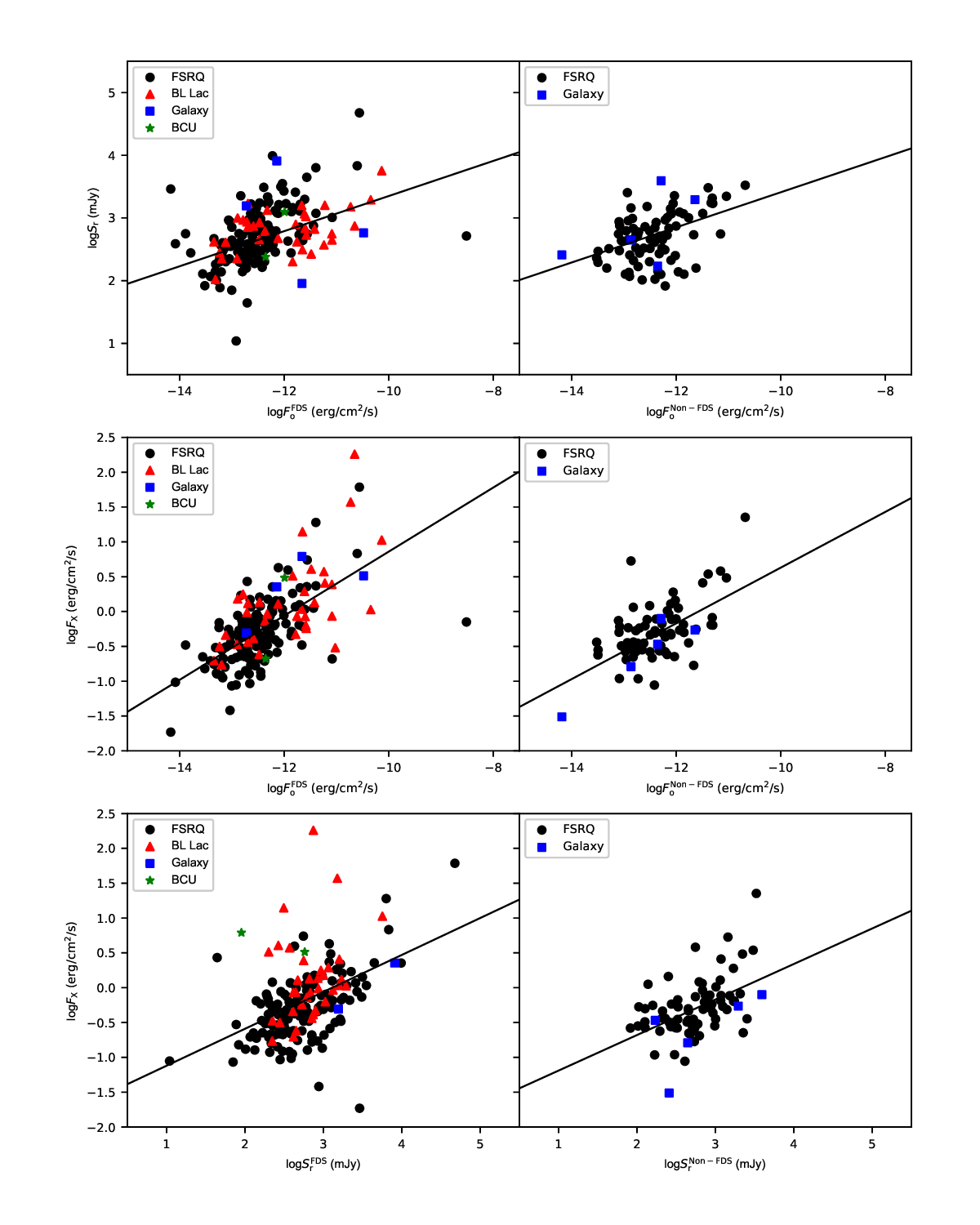}
  \caption{Plots for optical magnitude (log$F_{\rm o}$), radio flux density (log$S_{\rm r}$), X-ray flux (log$F_{\rm x}$), Left-hand panel for FDS sources, right-hand panel for non-FDS sources. Upper panel for radio against optical, middle panel for X-ray against optical and lower panel for X-ray against radio. Dot stands for FSRQs, triangle for BL Lacs and square for galaxies. The solid lines stand for best linear fitting results.}
  \label{F-F}
 \end{figure*}

The multi-wavelength (radio, optical and X-ray) data are from the BZCAT,
 the optical magnitude is made galactic extinction correction and then transferred into optical flux density.
 The optical,
  radio and X-ray flux densities are also K-corrected by $(1+z)^{\alpha-1}$,
  where $\alpha$ ($F_{\nu} \propto \nu^{-\alpha}$) is the spectral index in the given band.
For the spectral indexes,
 we adopt $\alpha_{\rm r}=0$ for radio band (\citealt{Donato2001}, \citealt{Abdo2010a}),
  while for optical band,
    $\alpha_{\rm o}=0.5$ for BLs and $\alpha_{\rm o}=1$ for the rest of the sources as did by
    \citet{Donato2001},
    $\alpha_{\rm X}=0.78$ for FSRQs,
    $\alpha_{\rm X}=1.30$ for BLs and
   $\alpha_{\rm X}=1.05$ for BCUs from \citet{Fan2016}.
   For any two bands, we have \\
$${\rm log}S_{\rm r} = (0.28 \pm 0.04) {\rm log}F_{\rm o}+ (6.15 \pm 0.49), ~r=0.51~{\rm and}~p=5.9 \times 10^{-14},$$
$${\rm log}F_{\rm x} = (0.46 \pm 0.04) {\rm log}F_{\rm o}+ (5.46 \pm 0.51), ~r=0.64~{\rm and}~p=6.8 \times 10^{-22}, ~{\rm and}~$$
$${\rm log}F_{\rm x} = (0.53 \pm 0.08) {\rm log}S_{\rm r}- (1.65 \pm 0.22), ~r=0.46~{\rm and}~p=1.0 \times 10^{-10}$$
for FDSs, and \\
$${\rm log}S_{\rm r} = (0.28 \pm 0.07) {\rm log}F_{\rm o}+ (6.21 \pm 0.82), ~r=0.38~{\rm and}~p=4.0 \times 10^{-4},$$
$${\rm log}F_{\rm x} = (0.40 \pm 0.06) {\rm log}F_{\rm o}+ (4.63 \pm 0.78), ~r=0.47~{\rm and}~p=1.8 \times 10^{-5}, ~{\rm and}~$$
$${\rm log}F_{\rm x} = (0.51 \pm 0.11) {\rm log}S_{\rm r}- (1.70 \pm 0.30), ~r=0.50~{\rm and}~p=3.7 \times 10^{-6}$$
for non-FDSs. These corresponding results are shown in Figure \ref{F-F}.

For the subclasses of BL Lacs and FSRQs,
 their correlations are shown in Table \ref{Tab-F-F} for FDSs and non-FDSs.

\begin{table*}
\centering
\caption{Flux density-Flux density correlation analysis results}
\label{Tab-F-F}
\begin{tabular}{|c|c|c|c|c|c|c|c|}
 \hline
  ~Type~& ~band~ & ~ Sample ~ & $~ a+ \Delta a ~$ & $~ b+\Delta b ~$ & $~ N ~$ & $~ r ~$ & $~ p ~$ \\   \hline

  & & whole    & $0.28 \pm 0.04$ & $6.15 \pm 0.49$ & 186 & 0.51 & $5.9 \times 10^{-14}$ \\   \cline{3-8}
  & ${\rm log}S_{\rm r} ~vs ~ {\rm log}F_{\rm o}$ & BL Lac       & $0.19 \pm 0.06$ & $5.08 \pm 0.70$ &  36  & 0.49 & $2.5 \times 10^{-3}$   \\   \cline{3-8}
  & & FSRQ & $0.34 \pm 0.05$ & $7.00 \pm 0.64$ & 144 & 0.49 & $3.3 \times 10^{-10}$  \\   \cline{2-8}

  & & whole & $0.46 \pm 0.04$ & $5.46 \pm 0.51$ & 180 & 0.64 & $6.8 \times 10^{-22}$ \\   \cline{3-8}
 FDS &  ${\rm log}F_{\rm X} ~vs~ {\rm log}F_{\rm o}$ & BL Lac & $0.45 \pm 0.10$ & $5.51 \pm 1.19$ & 36 & 0.61 & $6.7 \times 10^{-5}$ \\   \cline{3-8}
  & & FSRQ & $0.41 \pm 0.05$ & $4.87 \pm 0.60$ & 139 & 0.59 & $1.4 \times 10^{-14}$ \\   \cline{2-8}

  & & whole & $0.53 \pm 0.08$ & $-1.65 \pm 0.22$ & 179 & 0.46 & $1.0 \times 10^{-10}$ \\   \cline{3-8}
  & ${\rm log}F_{\rm X} ~vs~ {\rm log}S_{\rm r}$ & BL Lac & $0.58 \pm 0.34$ & $-1.51 \pm 0.95$ & 35 & 0.29 & $9.4 \%$ \\   \cline{3-8}
  & & FSRQ & $0.52 \pm 0.07$ & $-1.72 \pm 0.20$ & 138 & 0.52 & $3.9 \times 10^{-11}$ \\   \hline

  & ${\rm log}S_{\rm r} ~vs~ {\rm log}F_{\rm o}$ & whole & $0.28 \pm 0.07$ & $6.21 \pm 0.82$ & 82 & 0.38 & $4.0 \times 10^{-4}$ \\   \cline{3-8}
  & & FSRQ & $0.28 \pm 0.07$ & $6.22 \pm 0.86$ & 77 & 0.43 & $1.0 \times 10^{-4}$ \\   \cline{2-8}

 non-FDS & ${\rm log}F_{\rm X} ~vs~ {\rm log}F_{\rm o}$  & whole & $0.40 \pm 0.06$ & $4.63 \pm 0.78$ & 76 & 0.47 & $1.8 \times 10^{-5}$ \\   \cline{3-8}
  & & FSRQ & $0.36 \pm 0.07$ & $4.22 \pm 0.85$ & 71 & 0.54 & $1.2 \times 10^{-6}$ \\   \cline{2-8}

  & ${\rm log}F_{\rm X} ~vs~ {\rm log}S_{\rm r}$ & whole & $0.51 \pm 0.11$ & $-1.70 \pm 0.30$ & 76 & 0.50 & $3.7 \times 10^{-6}$ \\   \cline{3-8}
  & & FSRQ & $0.51 \pm 0.11$ & $-1.68 \pm 0.30$ & 71 & 0.49 & $1.5 \times 10^{-5}$ \\   \hline

\end{tabular}
\end{table*}

\section{Discussions}

 As a subclass of AGNs,
  blazar has many extreme observation properties,
   which may be attributed to a relativistic beaming effect.
  We compile proper motions, $\mu$ for 1216 components for 291 sources,
  including 189 FDS and 102 non-FDS sources.
  Then we collect and calculate the corresponding apparent velocity ($\beta_{\rm app}$),
  get their multi-wavelength data,
  and Doppler factors from available references (BZACT, MOJAVA,
  {\textit{Fermi}}-3rd catalogue, and so on),
  make comparisons between FDSs and non-FDSs,
  and then investigate some statistical correlations.

\subsection{Averaged Values for Superluminal Motions}

The averaged maximum values of proper motion for FDSs and non-FDSs are
 $\langle \mu^{\rm max}_{\rm FDS} \rangle=0.361 \pm 0.037$ and $\langle \mu^{\rm max}_{\rm non-FDS} \rangle=0.224 \pm 0.027$;
 while the averaged mean values of proper motion are
  $\langle \mu^{\rm mean}_{\rm FDS} \rangle=0.233 \pm 0.021$ and
  $\langle \mu^{\rm mean}_{\rm non-FDS} \rangle=0.158 \pm 0.015$ respectively.
  The averaged maximum values of apparent velocity for FDS and non-FDS sources are
  $\langle \beta^{\rm max}_{\rm FDS}\rangle=12.36 \pm 1.64$ and
  $\langle \beta^{\rm max}_{\rm non-FDS} \rangle=8.75 \pm 1.32$;
  the averaged mean values are
   $\langle \beta^{\rm mean}_{\rm FDS} \rangle=8.17 \pm 0.94$ and
   $\langle \beta^{\rm mean}_{\rm non-FDS} \rangle=5.99 \pm 0.78$.

 Based on MOJAVE 1.5 Jy flux density-limited samples,
 \citet{Lister2016} get a less than 0.02 \% probability that the LAT and non-LAT sub-samples
 come from the same parent distribution according to K-S tests.
When a K-S test is adapted to these distributions of FDSs and non-FDSs,
 we find that probability for the distributions of FDSs and non-FDSs
 to be from the same distribution is less than $1.6 \times 10^{-4}$ as shown in
  Figure \ref{mu-dis}-\ref{beta-dis},
  suggesting FDSs and non-FDSs should be from two different distributions.
 Our result confirms the result by
 \citet{Lister2016}.
We can say that FDSs have a larger proper motion and apparent velocity than do non-FDSs.

These results are consistent with other  results.
 \citet{Jorstad2001} indicated that the sources with $\gamma$-ray emission
show greater apparent velocities ($\beta_{app}$) than these sources without $\gamma$-ray emissions.
 \citet{Lister2009}, \citet{Lister2016} and
 \citet{Piner2012} also confirmed this result,
 which means that the $\gamma$-ray sources are highly beamed.
 Our result based on the largest superluminal sample also confirms their results.

\subsection{Correlations}

\citet{VC1994} collected 66 sources with proper motions,
 they showed that the proper motion decreases with increasing redshift (see also,
 \citealt{Cohen2005}).
 In our previous work
(\citealt{ZF2008}),
 we collected a sample of 123 superluminal sources,
 investigated the correlation between proper motion and redshift,
 and got ${\rm log} \mu \sim {-0.28} {\rm log} z$. In this paper,
  we obtained ${\rm log} \mu^{\rm max} =-( 0.28 \pm 0.07 ) {\rm log} z -( 0.95 \pm 0.02 )$
   with $r = 0.22$ and a chance probability of $p = 1.2 \times 10^{-4}$ for the whole sample.
   The result indicates clearly that the proper motion decreases with the increasing redshift,
   which is consistent with the results from others' 
(\citealt{VC1994},
\citealt{Cohen2005} and
\citealt{ZF2008}).
 When we consider FDS and non-FDS separately,
  there is an anti-correlation with a slope of $-0.35$ for the FDSs,
   and an anti-correlation tendency with a slope of $-0.14$ for the non-FDSs.

 When we investigate linear correlations between apparent velocities (log$\beta_{\rm app}$) and redshift (log$z$),
 slopes are 0.36 and 0.59 are obtained from the log$\beta_{\rm app}$-log$z$ fitting for FDS and non-FDS
sources respectively,
   which means the apparent velocity increases with redshift.
 Obviously, for non-FDS,
  the positive correlation is much better which with a chance probability is $p = 1.6 \times 10^{-7}$.
  \citet{Lister2009} plotted maximum apparent velocity against redshift for their sample and
  showed that the maximum superluminal velocity increases with redshift.
In the MOJAVE survey,
 the minimum detectable luminosity rises sharply with redshift,
 creating a classical Malmquist bias and the high redshift sources have higher apparent luminosities,
  which they achieved primarily via Doppler boosting
  (\citealt{Lister2009}).
Our result is consistent with theirs.

\citet{Kellermann2007} quoted their result of apparent velocity and luminosity and
indicated that there are no low luminosity sources with fast motions.
The high luminosity sources show a wider range of apparent velocity.
In Figure \ref{beta-L}, we also find such a tendency that the brighter $\gamma$-ray sources
 show a higher apparent velocity.
 It is due to a beaming effect because a source with a higher velocity  suggests
  a corresponding higher Doppler factor,
  and a higher Doppler boosting results in a higher luminosity.

From Fig. \ref{beta-L},
we can see a tendency for higher apparent velocity source to have higher $\gamma$-ray luminosity.
However, we can also see that some luminous $\gamma$-ray source have also low apparent
velocity. Actually, there is an envelope between apparent velocity and $\gamma$-ray luminosity in Figure \ref{beta-L},
 which is similar to that seen in the Caltech-Jodrell Bank Flat Spectrum (CJF) survey
(\citealt{Vermeulen1995}),
 the 2 cm Survey
 (\citealt{Kellermann2004}),
 and the MOJAVE survey
 (\citealt{Cohen2007};
 \citealt{Lister2009};
 \citealt{Piner2012}).
This upper envelope is not due to selection effects, although its precise physical origin is unclear. \citet{Lister2009} speculated that such an envelope may arise because of an intrinsic relation between jet speed and luminosity in the parent population.

\subsection{Flux-Flux Correlations}

The mutual correlations between fluxes were investigated in literature.
\citet{Fan1994} compiled 52 X-ray selected BL Lacs to study the mutual correlations among radio, X-ray and optical data and found closely mutual correlations.
\citet{Dondi1995} studied correlations between $\bar{L}_{\rm \gamma}$ versus $\bar{L}_{\rm r}$, $\bar{L}_{\rm o}$ and $\bar{L}_{\rm X}$, for a sample of quasars detected by EGRET, obtained positive correlations, and found that $\bar{L}_{\rm \gamma}$ correlates closer with $\bar{L}_{\rm r}$ than with $\bar{L}_{\rm o}$ or $\bar{L}_{\rm X}$.
In the present work, we investigate mutual correlations for radio, optical and X-ray fluxes for both FDS and non-FDS sources. For our flux-flux correlation analysis, we can see positive correlations and that there is no significant difference in slopes and intercepts between FDSs and non-FDSs.
In 2010, \citeauthor{Abdo2010a} used quasi-simultaneous data to calculate the
SEDs for a sample of 48 LAT Bright AGN sample sources (LBAS).
From their multwavelength data, we get following corresponding results,
$${\rm log}S_{\rm r}=(0.13 \pm 0.06) {\rm log}F_{\rm o}+(4.37 \pm 0.71), ~r=0.37~{\rm and}~p=1.2 \%, $$
$${\rm log}F_{\rm X}=(0.58 \pm 0.09) {\rm log}F_{\rm o}+(7.10 \pm 1.06), ~r=0.80~{\rm and}~p=3.1 \times 10^{-7}, and $$
$${\rm log}F_{\rm X}=(0.68 \pm 0.23) {\rm log}S_{\rm r}-(1.87 \pm 0.69), ~r=0.36~{\rm and}~p=1.7 \%. $$
 We can see that the results based on the quasi-simultaneous data are similar to our results.

The emissions from AGNs are mainly from the jet,
however there are contaminations from the host galaxy (especially for BL Lac objects)
as well as the big blue bump (especially for FSRQs) in the optical band,
together with the contribution from the accretion system in the X-ray domain.
Since the sample considered here is all superluminal, their jet emission should
be strongly boosted, which make the contaminations in the optical and X-ray bands
be relatively small.

\begin{figure}
 \centering
 \includegraphics[width=6in]{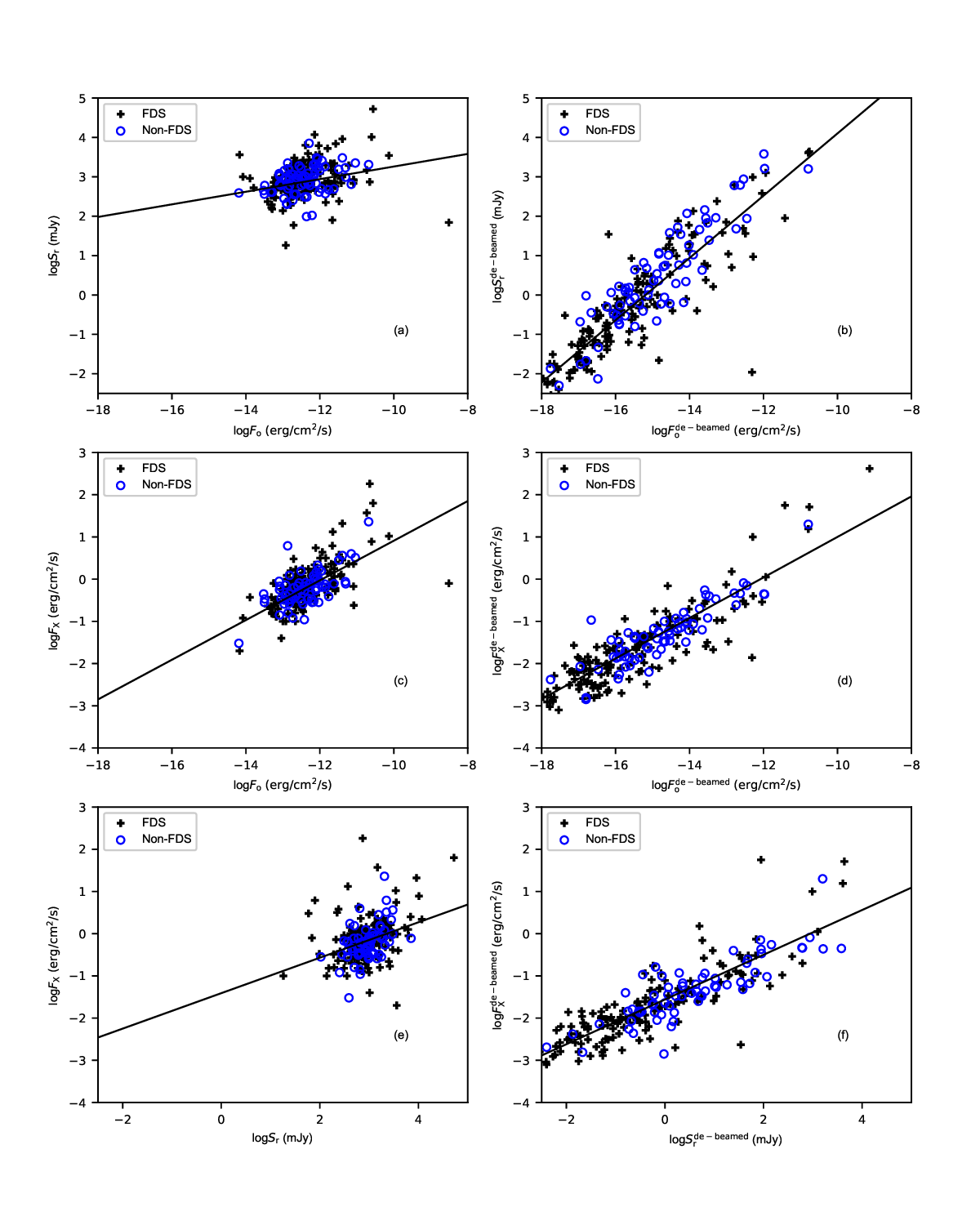}
  \caption{Mutual correlations between two bands flux densities. Left-hand panel for the observed data, right-hand panel for the removed beaming effect data. The upper panel, (a): radio flux density against optical flux density, (b): de-beamed radio flux density against de-beamed optical flux density; the middle panel, (c): X-ray flux density against optical flux density, (d): de-beamed X-ray flux density against de-beamed optical flux density; and the lower panel, (e): X-ray flux density against radio flux density, (f): de-beamed X-ray flux density against de-beamed radio flux density. The black plus for FDS sources and the blue circle for non-FDS sources and solid lines stand for best linear fitting results.}
  \label{de-F-F}
\end{figure}

\citet{Ghisellini1989} proposed that the bulk velocity of the plasma increases
with distance from the core and synchrotron X-rays are weakly beamed,
while optical and radio emissions are more strongly beamed.
\citet{Fan1993} proposed an empirical frequency dependent Doppler factor:
$\delta_{\rm \nu}=\delta_{\rm o}^{1+1/8~ {\rm log} ({\nu}_{\rm o}/{\nu})}$,
where $\delta_{\rm o}$ is the optical Doppler factor,
then $\delta_{\rm X} \sim \delta_{\rm o}^{0.5}$,
$\delta_{\rm r} \sim \delta_{\rm o}^{1.5}$,
$\delta_{\rm X}$ and $\delta_{\rm r}$ are the X-ray and radio Doppler factors
(\citealt{Fan1993}).
In a beaming model,
 the observed emission,
 $f^{\rm ob}$,
 is strongly boosted,
 namely, $f^{\rm ob} = \delta^{p} f^{\rm in}$,
 here $f^{\rm in}$ is the intrinsic emission in the source frame,
 $\delta$ is the Doppler factor,
 $p = 3 + \alpha$ is for a moving compact source or
 $p = 2 + \alpha$ for a continuous jet
(\citealt{Lind1985}).
Here $ p = 2 + \alpha$ is used as we did before
(\citealt{Xiao2015}).
In the work, Doppler factors are available for 224 sources,
 which makes it possible for us to investigate the mutual correlations
 for the intrinsic (de-beamed) radio, optical and X-ray flux emissions.
 For the 224 sources, we
 have \\
$${\rm log}S_{\rm r}=(0.16 \pm 0.04) {\rm log}F_{\rm o}+(4.86 \pm 0.47), ~r=0.34~{\rm and}~p=1.5 \times 10^{-7}, $$
$${\rm log}F_{\rm X}=(0.47 \pm 0.04) {\rm log}F_{\rm o}+(5.61 \pm 0.47), ~r=0.60~{\rm and}~p=9.6 \times 10^{-23}, $$
$${\rm log}F_{\rm X}=(0.42 \pm 0.07) {\rm log}S_{\rm r}-(1.41 \pm 0.23), ~r=0.36~{\rm and}~p=5.4 \times 10^{-8}, $$
for the observed data; and \\
$${\rm log}S_{\rm r}^{\rm de-beamed}=(0.79 \pm 0.03) {\rm log}F_{\rm o}^{\rm de-beamed}+(12.02 \pm 0.42), ~r=0.88~{\rm and}~p= 2.2 \times 10^{-74}, $$
$${\rm log}F_{\rm X}^{\rm de-beamed}=(0.48 \pm 0.02) {\rm log}F_{\rm o}^{\rm de-beamed}+(5.80 \pm 0.28), ~r=0.86~{\rm and}~p= 7.4 \times 10^{-63}, $$
$${\rm log}F_{\rm X}^{\rm de-beamed}=(0.53 \pm 0.02) {\rm log}S_{\rm r}^{\rm de-beamed}-(1.56 \pm 0.03), ~r=0.85~{\rm and}~p= 4.8 \times 10^{-62}, $$
for the intrinsic data.
It is clear that the correlations between any two bands become much closer when the beaming effect is removed.
The result indicates that the beaming effect affects the observed broad band correlations.
The comparison results are shown in Figure \ref{de-F-F}.

\subsection{Basic parameters for Jets}

\subsubsection{Doppler Factor, Lorentz Factor and Viewing Angle}

Doppler factors are important but not easy to estimate, although several methods were proposed. In the present work, we collect Doppler factor for 229 sources (151 FDS and 78 non-FDS) and list them in Column 4 in Tables \ref{Tab-FDS-simp} and \ref{Tab-non-FDS-simp}.

In a beaming model, the Lorentz factor ($\Gamma$) and viewing angle ($\phi$) can be obtained from $\delta$ and $\beta_{\rm app}$:

$$\Gamma =\frac{\beta_{\rm app}^{2}+\delta^{2}+1}{2\delta},
{\rm tan}\phi =\frac{2\beta_{\rm app}}{\beta_{\rm app}^{2}+\delta^{2}-1}.$$

From the Doppler factors in Tables \ref{Tab-FDS-simp} and \ref{Tab-non-FDS-simp}, we have,
$$\langle \delta^{\rm FDS} \rangle= 17.23 \pm 12.54 ~{\rm and}~ \langle \delta^{\rm non-FDS} \rangle= 9.48 \pm 8.86, $$
a K-S test result shows that
the probability for the two distributions to be from the same one
 is $7.5 \times 10^{-7}$.

For the sources with available Doppler factors, we can calculate $\Gamma$ and $\phi$, and get  their
mean values:
$$\langle \Gamma^{\rm FDS} \rangle= 21.41 \pm 21.59 ~{\rm and}~ \langle \Gamma^{\rm non-FDS} \rangle= 13.57 \pm 12.75, $$
 and
$$\langle \phi^{\rm FDS} \rangle= 5.64^{\circ} \pm 9.69^{\circ} ~{\rm and}~ \langle \phi^{\rm non-FDS} \rangle= 8.94^{\circ} \pm 7.77^{\circ},$$
with chance probabilities being $p = 5.2 \times 10^{-5}$ and $p = 1.5 \times 10^{-7}$ respectively,
which show significant difference in Lorentz factor and viewing angle between FDS and non-FDS sources.

\citet{Savolainen2010} obtained that for the photons arriving to us at an angle $\phi$, the jet flow is at an angle $\phi_{co}$ in the co-moving frame:
\begin{equation}
\phi_{\rm co} = {\rm arccos}(\frac{{\rm cos} \phi-\beta}{1-\beta {\rm cos} \phi})
\end{equation}
then, from the obtained $\Gamma$ and $\phi$, we can get $\phi_{co}$, and their mean values are
$$\langle \phi_{\rm co}^{\rm FDS} \rangle= 82.76^{\circ} \pm 46.83^{\circ} ~{\rm and}~ \langle \phi_{\rm co}^{\rm non-FDS} \rangle= 93.68^{\circ} \pm 43.21^{\circ}$$
with a K-S test result of $p = 15.8 \%$ suggesting no clear difference in co-moving viewing angle between FDSs and non-FDSs.

From K-S test results, we can see that there are significant differences in Doppler factor,
viewing angle and Lorentz factor between FDSs and non-FDSs.
FDSs show higher Doppler factors, higher Lorentz factor, and smaller viewing angle than do non-FDSs. The fact that FDS sources have larger Doppler factor also confirmed by \citet{Lister2015} who used the MOJAVE sample as well. However, the co-moving viewing angles show no clear difference between FDSs and non-FDSs. It means that FDS and non-FDS jets have similar cone in the comoving frame.

Above analysis results suggest that the difference between {\textit{Fermi}} detected superluminal sources
and {\textit{non-Fermi}} detected superluminal sources comes from their difference beaming effect
with FDSs being strongly beamed than non-FDSs. The superluminal source 0007+106 (III ZW2)
was not listed in the 3FGL, but it was classified as a $\gamma$-ray source by \citet{Liao2016}.
So, we propose  that the superluminal source is a $\gamma$-ray candidate and
the $\gamma$-ray source should be a superluminal source.
Our sample gives that $\langle \beta^{\rm max}_{\rm non-FDS} \rangle=8.75 \pm 1.32$ for non-FDS sources, and $\langle \beta^{\rm max}_{\rm FDS}\rangle=12.36 \pm 1.64$ for the FDS sources. If we take non-Fermi sources with $\beta^{\rm max}_{\rm non-FDS} > \langle \beta^{\rm max}_{\rm FDS}\rangle + 5\sigma$ as the candidate of $\gamma$-ray emitter, then there are 6 $\gamma$-ray emitting candidates, and they are 0153+744, 0208-512, 0536+145, 0552+398, 2223+210, 2351+456.

\subsubsection{$\gamma$-ray Luminosity and Viewing Angle}

\begin{figure}
 \centering
 \includegraphics[width=5in]{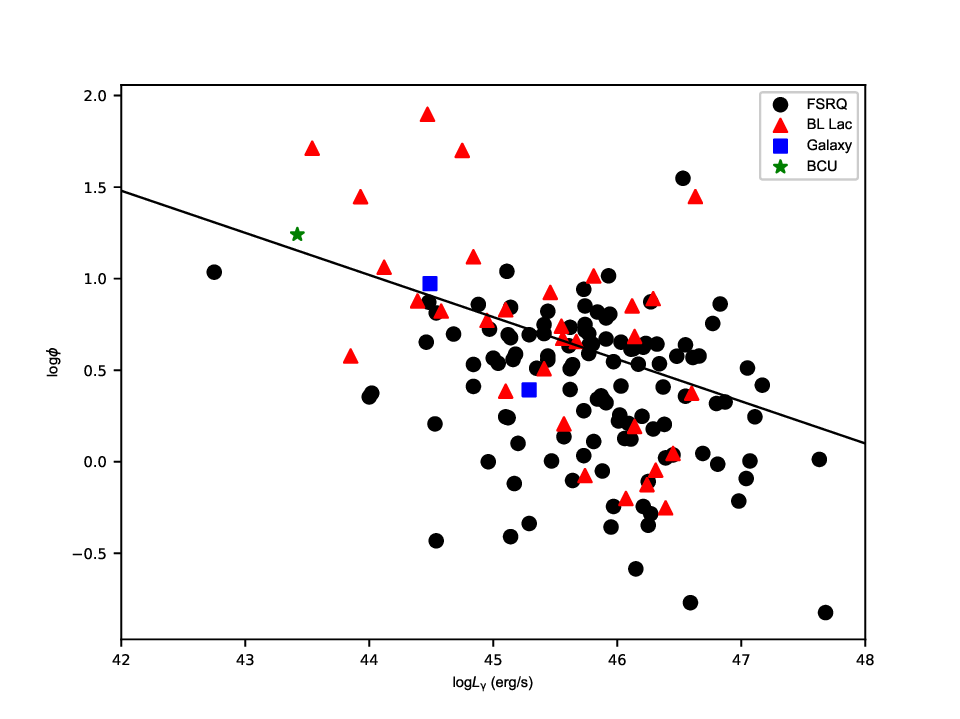}
  \caption{Correlation between observed $\gamma$-ray luminosity and viewing angle of FDS sources. The solid line shows the best linear fitting result of all the sources.}
  \label{phi-L}
\end{figure}

For $\gamma$-ray luminosity ($L_{\rm \gamma}$) and viewing angle ($\phi$), there is a significant anti-correlation,
$${\rm log}\phi=-(0.23 \pm 0.04) {\rm log}L_{\rm \gamma}+(11.14 \pm 1.93),$$
with $r=-0.38~{\rm and}~p= 2.2 \times 10^{-6}$. The best fitting result is shown in Figure \ref{phi-L} with a solid line. The results imply that the more luminous $\gamma$-ray sources have smaller viewing angles.

\begin{figure}
 \centering
 \includegraphics[width=5in]{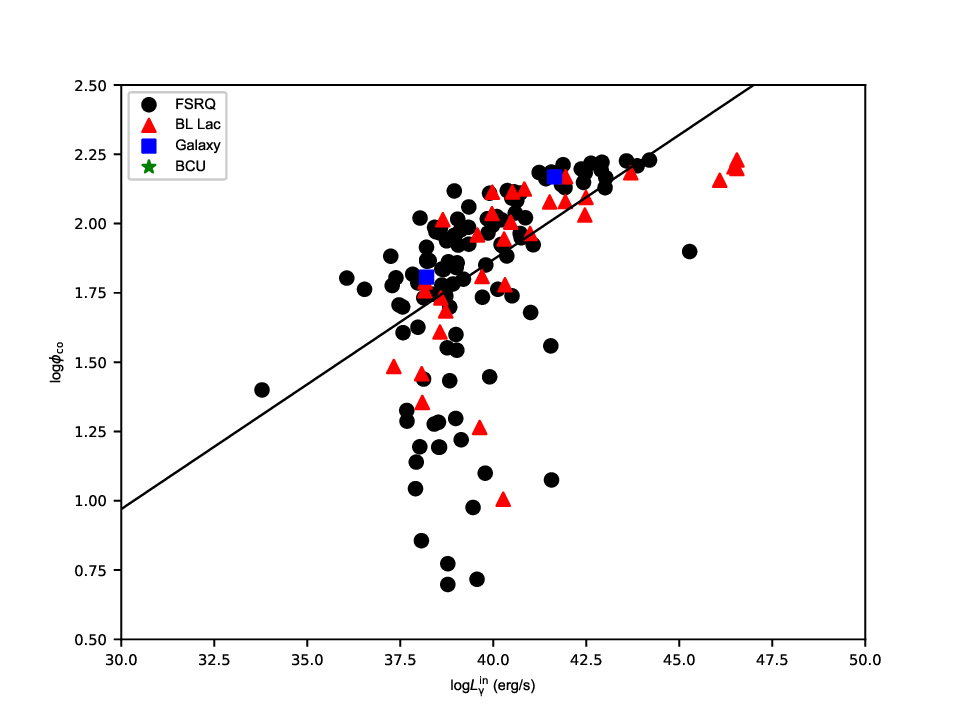}
  \caption{Correlation between intrinsic $\gamma$-ray luminosity and co-moving viewing angle of FDS sources. The solid line shows the best linear fitting result.}
  \label{phico-de-L}
\end{figure}

However, in a beaming model, the observed flux density, $f^{\rm ob}$ is correlated with the intrinsic flux density, $f^{\rm in}$, by $f^{\rm ob} = \delta^{p} f^{\rm in}$, where $p=3+\alpha$ for a discrete case and $p=2+\alpha$ for a continuous case. So, for the luminosity, we have $L^{\rm ob} =\delta^{4+\alpha} L^{\rm in}$ for a discrete case, or $L^{\rm ob} =\delta^{3+\alpha} L^{\rm in}$ for the continuous case. For sources with available $\delta$, we can get their intrinsic luminosity and co-moving viewing angle, which show \\
$${\rm log}\phi_{\rm co}=(0.09 \pm 0.01) {\rm log}L_{\rm \gamma}^{\rm in}-(1.73 \pm 0.48),$$
with $r=0.69~{\rm and}~p= 7.3 \times 10^{-22}$, the result is shown in Figure \ref{phico-de-L} with a solid line.

Obviously, there is a positive correlation between co-moving viewing angle and intrinsic luminosity,
 which suggests that the luminous intrinsic $\gamma$-ray luminosity corresponds to a wider co-moving viewing angle.
If the emission per solid angle in the comoving is similar for {\textit{Fermi-LAT}} sources,
then the larger co-moving angle corresponds to larger solid angle frame,
then the emission is stronger.

From Figures \ref{beta-L}-\ref{phico-de-L}, we can see that FDS and non-FDS sources are different. FDSs show higher proper motion ($\mu$), higher apparent velocity ($\beta_{\rm app}$), higher Doppler factor ($\delta$), higher Lorentz factor ($\Gamma$) and smaller viewing angle ($\phi$) than non-FDSs.
We can say that FDSs have stronger beaming effect than do non-FDSs.

\section{Conclusions}

The 291 superluminal sources had 1216 components.
We collected their multi-wavelength data and Doppler factors and calculated the
apparent velocity for each component,
the Lorentz factor and viewing angle (co-moving angle) for those with an available Doppler factor,
calculate the $\gamma$-ray luminosity for the 3GFL sources.
 Subsequently, we investigated the relationships between FDS and non-FDS sources.
 Our main conclusions are summarized as follows:

1. FDS sources show higher proper motion, apparent velocity, Doppler factor, Lorentz factor and smaller viewing angles than non-FDS sources. For superluminal sources, FDSs are more beamed.

2. The intrinsic (de-beamed) fluxes show much closer mutual correlations among radio, optical and X-ray bands than the observed data.

3. A higher apparent velocity source has the tendency to exhibit higher $\gamma$-ray luminosity,
 and the superluminal sources may be the $\gamma$-ray emission candidates;
  moreover $\gamma$-ray sources with no known superluminal velocity can be superluminal candidates.

4. The $\gamma$-ray brighter source shows a smaller viewing angle,
 which suggest a strong Doppler effect.
   High energetic $\gamma$-ray emissions per solid angle are probably similar in the co-moving frame;
   thus  the de-beamed $\gamma$-ray luminosity is positively correlated with the co-moving viewing angle.

\begin{acknowledgements}
This work is partially supported by the National Natural Science
Foundation of  China (NSFC 11733001, NSFC U1531245, NSFC 10633010, NSFC 11173009,NSFC 11403006),  Natural Science Foundation of Guangdong Province
(2017A030313011), supports for Astrophysics  Key Subjects of Guangdong Province and Guangzhou City, and Science and Technology Program of Guangzhou (201707010401). We also thank the MOJAVE team and Purdue University for use the precious data of kinematic details. We thank the referees for the useful comments and constructive suggestions.

\end{acknowledgements}

\appendix
\section{Appendix: Full version of Tables}
\startlongtable
\begin{longrotatetable}
\begin{deluxetable}{c|c|c|c|c|c|c|c|c|c|c|c|c|c|c}
\tabletypesize{\tiny}
 \tablecaption{Full table of Table 1}
  \tablewidth{0pt}
 \tablehead{
  \colhead{ FGL name }&
  \colhead{ Class } &
  \colhead{ redshift } &
  \colhead{ $\delta_R$  } &
  \colhead{ Ref } &
  \colhead{ $m_o$  } &
  \colhead{ Ext  }&
  \colhead{ S$_R$  }&
  \colhead{ S$_X$  }&
  \colhead{ $\Gamma_{\gamma}$  }&
  \colhead{ F$\gamma$  }&
  \colhead{ $\mu$ }&
  \colhead{ comp }&
  \colhead{ Ref } &
  \colhead{ $\beta$ }\\
  \colhead{ Other name }&
  \colhead{ }&
  \colhead{ }&
  \colhead{ }&
  \colhead{ }&
  \colhead{ $magnitude $}&
  \colhead{ }&
  \colhead{ $mJy$ }&
  \colhead{ $1 \times 10^{-12} ~cgs$ }&
  \colhead{ }&
  \colhead{ $ph/cm^{2}/s$ }&
  \colhead{ $\mu as/yr$ }&
  \colhead{ }&
  \colhead{ }&
  \colhead{}\\
  \colhead{(1) }&
  \colhead{(2) }&
  \colhead{(3) }&
  \colhead{(4) }&
  \colhead{(5) }&
  \colhead{(6) }&
  \colhead{(7) }&
  \colhead{(8) }&
  \colhead{(9) }&
  \colhead{(10) }&
  \colhead{(11) }&
  \colhead{(12) }&
  \colhead{(13) }&
  \colhead{(14) }&
  \colhead{(15) }
    }
 \startdata
0007+106 	&	G	&	0.089	&	2.51	&	L18	&	15.8	&	0.227	&	98	&	6.14	&		&		&	204 $\pm$ 12	&	1	&	MOJAVE	&	1.197$\pm$0.069	\\	
III ZW 2	&		&		&		&		&		&		&		&		&		&		&	269 $\pm$ 50	&	4	&	MOJAVE	&	1.58$\pm$0.29	\\	\hline
1FGL J1159.4-2149	&	F	&	0.927	&		&		&	17.8	&	0.104	&	386	&	0	&		&		&	9.3 $\pm$ 3.8	&	1	&	MOJAVE	&	0.46$\pm$0.19	\\	
1157-215 	&		&		&		&		&		&		&		&		&		&		&	83 $\pm$ 22	&	2	&	MOJAVE	&	4.1$\pm$1.1	\\	
	&		&		&		&		&		&		&		&		&		&		&	26.2 $\pm$ 8.0	&	3	&	MOJAVE	&	1.30$\pm$0.40	\\	 \hline
1FGL J1245.8-0632	&	F	&	1.286	&		&		&	19.6	&	0.068	&	551	&	0	&		&		&	358 $\pm$ 53	&	1	&	MOJAVE	&	22.5$\pm$3.4	\\	
1243-072 	&		&		&		&		&		&		&		&		&		&		&	17.6 $\pm$ 8.9	&	3	&	MOJAVE	&	1.11$\pm$0.56	\\	
	&		&		&		&		&		&		&		&		&		&		&	40 $\pm$ 16	&	4	&	MOJAVE	&	2.51$\pm$1.00	\\
\enddata
\label{Tab-FDS}
\tablecomments{Note to the Table: B08: \citet{Britzen2008}; F09: \citet{Fan2009}; GPC99: \citet{GPC1999}; H09: \citet{Hovatta2009}; JM01: \citet{Jorstad2001}; L18: \citet{Liodakis2018}; PBE06: \citet{PBE06}; S10: \citet{Savolainen2010}; VC94: \citet{VC1994}; X95:\citet{Xu1995}; CJF: Caltech-Jodrell Bank Flat-spectrum sample: http://www.mpifr-bonn.mpg.de/staff/sbritzen/cjf.html; MOJAVE: Monitoring Of Jets in Active galactic nuclei with VLBA Experiments: http://www.physics.purdue.edu/MOJAVE/}
\end{deluxetable}
\end{longrotatetable}

\begin{longrotatetable}
\begin{deluxetable}{c|c|c|c|c|c|c|c|c|c|c|c|c}
\tabletypesize{\tiny}
  \tablecaption{Full table of Table 2}
  \tablewidth{0pt}
 \tablehead{
  \colhead{ FGL name }&
  \colhead{ Class } &
  \colhead{ redshift } &
  \colhead{ $\delta_R$  } &
  \colhead{ Ref } &
  \colhead{ $m_o$  } &
  \colhead{ Ext  }&
  \colhead{ S$_R$  }&
  \colhead{ S$_X$  }&
  \colhead{ $\mu$  }&
  \colhead{ comp  }&
  \colhead{ Ref  } &
  \colhead{ $\beta$  }\\
  \colhead{ Other name }&
  \colhead{ }&
  \colhead{ }&
  \colhead{ }&
  \colhead{ }&
  \colhead{ $magnitude$ }&
  \colhead{ }&
  \colhead{ $mJy$ }&
  \colhead{ $1 \times 10^{-12} ~cgs$ }&
  \colhead{$\mu as/yr$ }&
  \colhead{ }&
  \colhead{ }&
  \colhead{}\\
  \colhead{(1) }&
  \colhead{(2) }&
  \colhead{(3) }&
  \colhead{(4) }&
  \colhead{(5) }&
  \colhead{(6) }&
  \colhead{(7) }&
  \colhead{(8) }&
  \colhead{(9) }&
  \colhead{(10) }&
  \colhead{(11) }&
  \colhead{(12) }&
  \colhead{(13) }
    }
\startdata
0003-066	&	B	&	0.347	&	6.96	&	L18	&	17.9	&	0.087	&	2051	&	0.82	&	191 $\pm$ 15	&	2	&	MOJAVE	&	4.09$\pm$0.33	\\	
	&		&		&		&		&		&		&		&		&	250 $\pm$ 39	&	3	&	MOJAVE	&	5.36$\pm$0.83	\\	
	&		&		&		&		&		&		&		&		&	50.4 $\pm$ 5.3	&	4a 	&	MOJAVE	&	1.08$\pm$0.11	\\	
	&		&		&		&		&		&		&		&		&	100 $\pm$ 16	&	5	&	MOJAVE	&	2.15$\pm$0.35	\\	
	&		&		&		&		&		&		&		&		&	54 $\pm$ 11	&	6a 	&	MOJAVE	&	1.16$\pm$0.24	\\	
	&		&		&		&		&		&		&		&		&	330.4 $\pm$ 9.8	&	8a 	&	MOJAVE	&	7.08$\pm$0.21	\\	
	&		&		&		&		&		&		&		&		&	287 $\pm$ 25	&	9	&	MOJAVE	&	6.14$\pm$0.53	\\	
	&		&		&		&		&		&		&		&		&	116 $\pm$ 23	&	14	&	MOJAVE	&	2.48$\pm$0.50	\\
\enddata
\label{Tab-non-FDS}
\tablecomments{Note to the Table: B08: \citet{Britzen2008}; F09: \citet{Fan2009}; Gio99a: \citet{Giovannini1999}; Frey2014: \citet{Frey2014}; H09: \citet{Hovatta2009}; L18: \citet{Liodakis2018}; Marc85: \citet{Marcaide1985}; SHW98: \citet{Shen1998}; VC94: \citet{VC1994}; CJF: Caltech-Jodrell Bank Flat-spectrum sample: http://www.mpifr-bonn. mpg.de/staff/sbritzen/cjf.html; MOJAVE: Monitoring Of Jets in Active galactic nuclei with VLBA Experiments: http://www.physics.purdue.edu/MOJAVE/}
\end{deluxetable}
\end{longrotatetable}


\begin{thebibliography}{}

\bibitem[Abdo et al.(2010a)] {Abdo2010a} A.-A. Abdo, M. Ackermann, I. Agudo, et al., ApJ, 716, 30 (2010a).

\bibitem[Abdo et al.(2010b)] {Abdo2010b} A.-A. Abdo, M. Ackermann, M. Ajello, et al., ApJ, 710, 810 (2010b).

\bibitem[Pushkarev et al.(2010)] {Pushkarev2010} A.-B. Pushkarev, Y.-Y. Kovalev, M.-L. Lister, ApJ, 722L, 7 (2010).

\bibitem[L\"ahteenim\"aki \& Valtaoja.(1999)] {Lahteenimaki1999} A. L\"ahteenim\"aki \& E. Valtaoja, ApJ, 521, 493 (1999).

\bibitem[Piner et al.(2006)] {PBE06} B.-G. Piner, D. Bhattarai, P.-G. Edwards, et al., ApJ, 640, 196 (2006).

\bibitem[Piner et al.(2012)] {Piner2012} B.-G. Piner, A.-B. Pushkarev, Y.-Y. Kovalev, C.-J. Marvin, et al., ApJ, 758, 84 (2012).

\bibitem[Lin \& Fan.(2018)] {Lin2018} C. Lin \& J.-H. Fan, 2018, RAA, 18, 120 (2018).

\bibitem[von Montigny et al.(1995)] {vonMontigny1995} C. von Montigny, D.-L. Bertsch, et al., ApJ, 440, 525 (1995).

\bibitem[Gabuzda et al.(1999)] {GPC1999} D.-C. Gabuzda, A.-B. Pushkarev \& T.-V. Cawthorne, MNRAS, 307, 725 (1999).

\bibitem[Donato et al.(2001)] {Donato2001} D. Donato, G. Ghisellini, G. Tagliaferri and G. Fossati, A\&A, 375, 739 (2001).


\bibitem[Nieppola et al.(2006)] {Nieppola2006} E. Nieppola, M. Tornikoski \& E. Valtaoja, A\&A, 445, 441 (2006).

\bibitem[Acero et al.(2015)] {Acero2015} F. Acero, M. Ackermann, M. Ajello, A. Albert, et al., ApJS, 218, 23 (2015).

\bibitem[Massaro et al.(2013a)] {Massaro2013a} F. Massaro, M. Giroletti, A. Paggi, et al., ApJS, 207, 4 (2013a).

\bibitem[Massaro et al. (2013b)] {Massaro2013b} F. Massaro, R. D' Abrusco, M. Giroletti, et al., ApJS, 208, 15 (2013b).

\bibitem[Ghisellini et al.(1989)] {Ghisellini1989} G. Ghisellini, P. Padovani and L. Maraschi, ApJ, 340, 181 (1989).

\bibitem[Ghisellini et al.(1993)] {Ghisellini1993} G. Ghisellini, P. Padovani, A. Celotti, L. Maraschi, ApJ, 407, 65 (1993).


\bibitem[Ghisellini et al.(2014)] {Ghisellini2014} G. Ghisellini, F. Tavecchio, L. Maraschi, A. Celotti, \& T. Sbarrato, Nature, 515, 376 (2014).

\bibitem[Giovannini et al.(1999)] {Giovannini1999} G. Giovannini, G.-B. Taylor, E. Arbizzani, et al., ApJ, 522, 101 (1999).

\bibitem[Giovannini et al.(2014)] {Giovannini2014} G. Giovannini, E. Liuzzo, B. Boccardi, M. Giroletti, IAUS, 304, 200 (2014).

\bibitem[Xie et al.(1997)] {Xie1997} G.-Z. Xie, Y.-H. Zhang, J.-H. Fan, ApJ, 477, 114 (1997).

\bibitem[Xiao et al.(2015)] {Xiao2015} H.-B. Xiao, J.-H. Fan, Z.-Y. Pei, et al., Ap\&SS, 359, 39 (2015).

\bibitem[Liodakis et al.(2018)] {Liodakis2018} I. Liodakis, T. Hovatta, D. Huppenkothen, S. Kiehlmann, et al., ApJ, 886, 2 (2018).

\bibitem[Chang et al.(2017)] {Chang2017} J. Chang, G. Ambrosi, et al., Astroparticle Physics, 95 ,6 (2017).

\bibitem[Fan et al.(1993)] {Fan1993} J.-H. Fan, G.-Z. Xie, J.-J. Li, ApJ, 415, 113 (1993).

\bibitem[Fan et al.(1994)] {Fan1994} J.-H. Fan, Z.-H. Wang, G.-Z. Xie, et al., A\&AS, 105, 415 (1994).

\bibitem[Fan et al.(1999)] {Fan1999} J.-H. Fan, R. Bacon, G.-Z. Xie, A\&AS, 136, 113 (1999).

\bibitem[Fan(2005)] {Fan2005} J.-H. Fan, A\&A, 436, 799 (2005).

\bibitem[Fan et al.(2009)] {Fan2009} J.-H. Fan, J.-H. Yang, J.-Y. Zhang, et al., PASJ, 61, 639 (2009).

\bibitem[Fan et al.(2013a)] {Fan2013a} J.-H. Fan, J.-H. Yang, Y. Liu, J.-Y. Zhang, RAA, 13, 259 (2013a).

\bibitem[Fan et al.(2013b)] {Fan2013b} J.-H. Fan, J.-H. Yang, J.-Y. Zhang, et al., PASJ, 65, 25 (2013b).

\bibitem[Fan \& Ji.(2014)] {FanJi2014} J.-H. Fan \& Z.-Y. Ji, IAUS, 304, 159 (2014).

\bibitem[Fan et al.(2014)] {Fan2014} J.-H. Fan, D. Bastieri, J.-H. Yang, et al., RAA, 14, 1135 (2014).

\bibitem[Fan et al.(2016)] {Fan2016} J.-H. Fan, J.-H. Yang, Y. Liu,  et al., ApJS, 226, 20 (2016).

\bibitem[Fan et al.(2017)] {Fan2017} J.-H. Fan, J.-H. Yang, H.-B. Xiao, C. Lin, et al., ApJL, 835, 38 (2017).

\bibitem[Yang et al.(2017)] {Yang2017} J.-H. Yang, J.-H. Fan, Y. Liu, Y.-L. Zhang, R.-S. Yang, M.-X. Tuo, J.-J. Nie, Y.-H. Yuan, Ap\&SS, 362, 219 (2017).

\bibitem[Yang et al.(2018a)] {Yang2018a} J.-H. Yang, J.-H. Fan, Y.-L. Zhang, R.-S. Yang, M.-X. Tuo, J.-J. Nie, AcASn, 59, 4 (2018a).

\bibitem[Yang et al.(2018b)] {Yang2018b} J.-H. Yang, J.-H. Fan, Y. Liu, Y.-L. Zhang, M.-X. Tuo, J.-J. Nie, Y.-H. Yuan, SCPMA, 5, 59511 (2018b).

\bibitem[Marcaide et al.(1985)] {Marcaide1985} J.-M. Marcaide, I.-I. Shapiro, B.-E. Corey, et al., A\&A, 142, 71 (1985).

\bibitem[Mattox et al.(1993)] {Mattox1993} J.-R. Mattox, D.-L. Bertsch, J. Chiang, et al., ApJ, 410, 609 (1993).

\bibitem[Zhang et al.(2012)] {Zhang2012} J. Zhang, E.-W. Liang, S.-N. Zhang \& J.-M. Bai, ApJ, 752, 157 (2012).

\bibitem[Kellermann et al.(2003)] {Kellermann2003} K.-L. Kellermann, M.-L. Lister, D.-C. Homan, ASPC, 299, 117 (2003).

\bibitem[Kellermann et al.(2004)] {Kellermann2004} K.-L. Kellermann, M.-L. Lister, D.-C. Homan, ApJ, 609, 539 (2004).

\bibitem[Kellermann et al.(2007)] {Kellermann2007} K.-L. Kellermann, Y.-Y. Kovalev, et al., Ap\&SS, 311, 231 (2007).

\bibitem[Lind \& Blandford.(1985)] {Lind1985} K.-R. Lind \& R.-D. Blandford, ApJ, 295, 358 (1985).

\bibitem[Cheng et al.(1999)] {Cheng1999} K.-S. Cheng, J.-H. Fan, L. Zhang, A\&A, 352, 32 (1999).

\bibitem[Cheng et al.(2000)] {Cheng2000} K.-S. Cheng, X. Zhang, L. Zhang, ApJ,  537, 80 (2000).

\bibitem[Chen et al.(2018)] {Chen2018} L. Chen, ApJS, 235, 39 (2018).

\bibitem[Dondi \& Ghisellini.(1995)] {Dondi1995} L. Dondi \& G. Ghisellini, MNRAS, 273, 583 (1995).

\bibitem[Zhang \& Fan.(2018)] {ZF2018} L.-X. Zhang \& J.-H. Fan, Ap\&SS, 363, 142 (2018).

\bibitem[Ackermann.(2015)] {Ackermann2015} M. Ackermann, M. Ajello, W.-B. Atwood, et al., ApJ, 810, 14 (2015).

\bibitem[Cohen et al.(2005)] {Cohen2005} M.-H. Cohen, M.-L. Lister, R.-C. Vermeulen, EAS, 15, 93 (2005).

\bibitem[Cohen et al.(2007)] {Cohen2007} M.-H. Cohen, M.-L. Lister, D.-C. Homan. et al., ApJ, 658, 232 (2007).

\bibitem[Giroletti et al.(2012)] {Giroletti2012} M. Giroletti, V. Pavlidou, A. Reimer, et al., AdSpR, 49, 1320 (2012).

\bibitem[Lister et al.(2009)] {Lister2009} M.-L. Lister, M.-H. Cohen, D.-C. Homan, et al., AJ, 138, 1874 (2009).

\bibitem[Lister et al.(2013)] {Lister2013} M.-L. Lister, M.-F. Aller, H.-D. Aller, D.-C. Homan, et al., AJ, 146, 120 (2013).

\bibitem[Lister et al.(2015)] {Lister2015} M.-L. Lister, M.-F. Aller, H.-D. Aller, T. Hovatta, W. Max Moerbeck, ApJL, 810, 9L (2015).

\bibitem[Lister et al.(2016)] {Lister2016} M.-L. Lister, M.-F. Aller, H.-D. Aller, D.-C. Homan, et al., AJ, 152, 12 (2016).

\bibitem[Liao et al.(2016)] {Liao2016} N.-H. Liao, Y.-L. Xin, X.-L. Fan, S.-S. Weng, et al., ApJS, 226, 17 (2016).

\bibitem[Nolan et al.(2012)] {Nolan2012}  P.-L. Nolan, A.-A. Abdo, M. Ackermann, et al., ApJS, 199, 31 (2012).

\bibitem[Padovani \& Giommi.(1995)] {Padovani1995} P. Padovani \& P. Giommi, ApJ, 444, 567 (1995).

\bibitem[Vermeulen \& Cohen.(1994)] {VC1994} R.-C. Vermeulen \& M.-H. Cohen, ApJ, 430, 467 (1994).

\bibitem[Vermeulen.(1995)] {Vermeulen1995} R.-C. Vermeulen, PNAS, 92, 11385 (1995).

\bibitem[Blandford \& K\"onigl. et al.(1979)] {Blandford1979} R.-D. Blandford, A. K\"onigl, ApJ, 1, 232 (1979).

\bibitem[Britzen et al.(2008)] {Britzen2008} S. Britzen, R.-M. Vermeulen, et al., A\&A, 484, 119 (2008).

\bibitem[Frey et al.(2014)] {Frey2014} S. Frey, Z. Paragi, J. O. Fogasy, L. I. Gurvits, MNRAS, 446, 2921 (2015).

\bibitem[Jorstad et al.(2001)] {Jorstad2001} S.-G. Jorstad, A.-P. Marscher, et al., ApJS, 134, 181 (2001).

\bibitem[Jorstad et al.(2005)] {Jorstad2005} S.-G. Jorstad, A.-P. Marscher, M.-L. Lister, et al., AJ, 130, 1418 (2005).

\bibitem[Jorstad et al.(2017)] {Jorstad2017} S.-G. Jorstad, A.-P. Marscher, D.-A. Morozova, et al., ApJ, 846, 98 (2017).

\bibitem[Arshakian et al.(2010)] {Arshakian2010} T.-G. Arshakian, J. Torrealba, V.-H. Chavushyan, et al., A\&A, 520, A62 (2010).

\bibitem[Hovatta et al.(2009)] {Hovatta2009} T. Hovatta, E. Valtaoja, M. Tornikorski, et al., A\&A, 496, 527 (2009).

\bibitem[Savolainen et al.(2010)] {Savolainen2010} T. Savolainen, D.-C. Homan, T. Hovatta, et al., A\&A, 512A, 24 (2010).

\bibitem[Xu et al.(1995)] {Xu1995} W. Xu, A.-C.-S. Readhead, T.-J. Pearson, ApJS, 99, 297 (1995).

\bibitem[Zhang \& Fan.(2019)] {ZF2019}Y.-T. Zhang \& J.-H. Fan, AcASn, (in press) (2019).

\bibitem[Zhang \& Fan.(2008)] {ZF2008} Y.-W. Zhang \&  J.-H. Fan, 2008, ChJAA, 4, 385 (2008).

\bibitem[Kovalev et al.(2009)] {Kovalev2009} Y.-Y. Kovalev, et al., 2009, ApJ, 707, 56 (2009).

\bibitem[Shen et al.(1998)] {Shen1998} Z.-Q. Shen, X.-Y. Hong, T.-S. Wan, ChA\&A, 22, 133 (1998).

\bibitem[Pei et al.(2016)] {Pei2016} Z.-Y. Pei, J.-H. Fan, et al., Ap\&SS, 361, 237 (2016).

\end{thebibliography}
\end{document}